\documentclass[manuscript,screen,nonacm]{acmart} 
\acmJournal{TOMM}
\usepackage{enumitem}
\usepackage{siunitx}
\usepackage{graphicx}
\usepackage{tikz}
\usepackage[caption=false,font=normalfont]{subfig}
\usepackage{makecell}
\usepackage{threeparttable}
\usepackage{bm}
\usepackage[ruled,linesnumbered,vlined]{algorithm2e}
\usepackage{standalone}
\usetikzlibrary{shapes.geometric, arrows.meta, positioning, fit, backgrounds, calc}
\usepackage[normalem]{ulem}
\useunder{\uline}{\ul}{}

\definecolor{objgreen}{RGB}{13, 92, 46}
\definecolor{subjblue}{RGB}{10, 25, 110}

\setlist[enumerate]{nosep}

\SetAlFnt{\small}
\SetAlCapFnt{\small}
\SetAlCapNameFnt{\small}
\SetAlCapHSkip{0pt}
\IncMargin{-\linewidth}

\AtBeginDocument{%
  }

\setcopyright{acmlicensed}
\copyrightyear{2018}
\acmYear{2018}
\acmDOI{XXXXXXX.XXXXXXX}
\acmConference[Conference acronym 'XX]{Make sure to enter the correct
  conference title from your rights confirmation email}{June 03--05,
  2018}{Woodstock, NY}
\acmISBN{978-1-4503-XXXX-X/2018/06}

\begin{document}

\title[NAVIQUAL-XR]{NAVIQUAL-XR: Finding the Sweet Spot of a Live Music Environment in Augmented Reality}

\author{Carl Timothy Tolentino}
\correspondingauthor
\email{carl.tolentino@ucd.ie}
\orcid{0009-0004-9809-3766}
\author{Andrew Hines}
\email{andrew.hines@ucd.ie}
\orcid{0000-0001-9636-2556}
\affiliation{%
  \institution{University College Dublin}
  \city{Dublin}
  \country{Ireland}
}


\begin{abstract}
Augmented Reality (AR) enables immersive spatial audio experiences that extend conventional stereo and head-tracked listening to six degrees-of-freedom (6-DoF) navigation. However, evaluating listener Quality of Experience (QoE) remains a key challenge due to the real-time and dynamic adaptations of the content in AR environments. To address this challenge, we propose NAVIQUAL-XR, an objective model to find the ``sweet spot'' in AR live music environments validated by subjective tests. We recruited 25 listeners who navigated 12 AR environments of various scene set ups under auditory and audio-visual modalities. We analyzed their navigation trajectories and recorded their final locations and directions as their subjective sweet spots. We then produced subjective maps that identified the location and direction with the optimal QoE in an AR environment. Furthermore, we benchmarked key objective metrics that predict these sweet spots through similarity metrics with the subjective maps. We found that localization quality derived from BINAQUAL, an objective metric designed for binaural audio, outperformed signal-to-noise ratio in closely predicting the sweet spot. This suggests that phase-based cues play a more important role on perceived QoE in AR environments than intensity-based cues. NAVIQUAL-XR provides new insights into listener behavior and preferences in AR environments, demonstrating the potential of objective metrics to predict the subjective experience, and guiding the design of future immersive audio systems. 
\end{abstract}




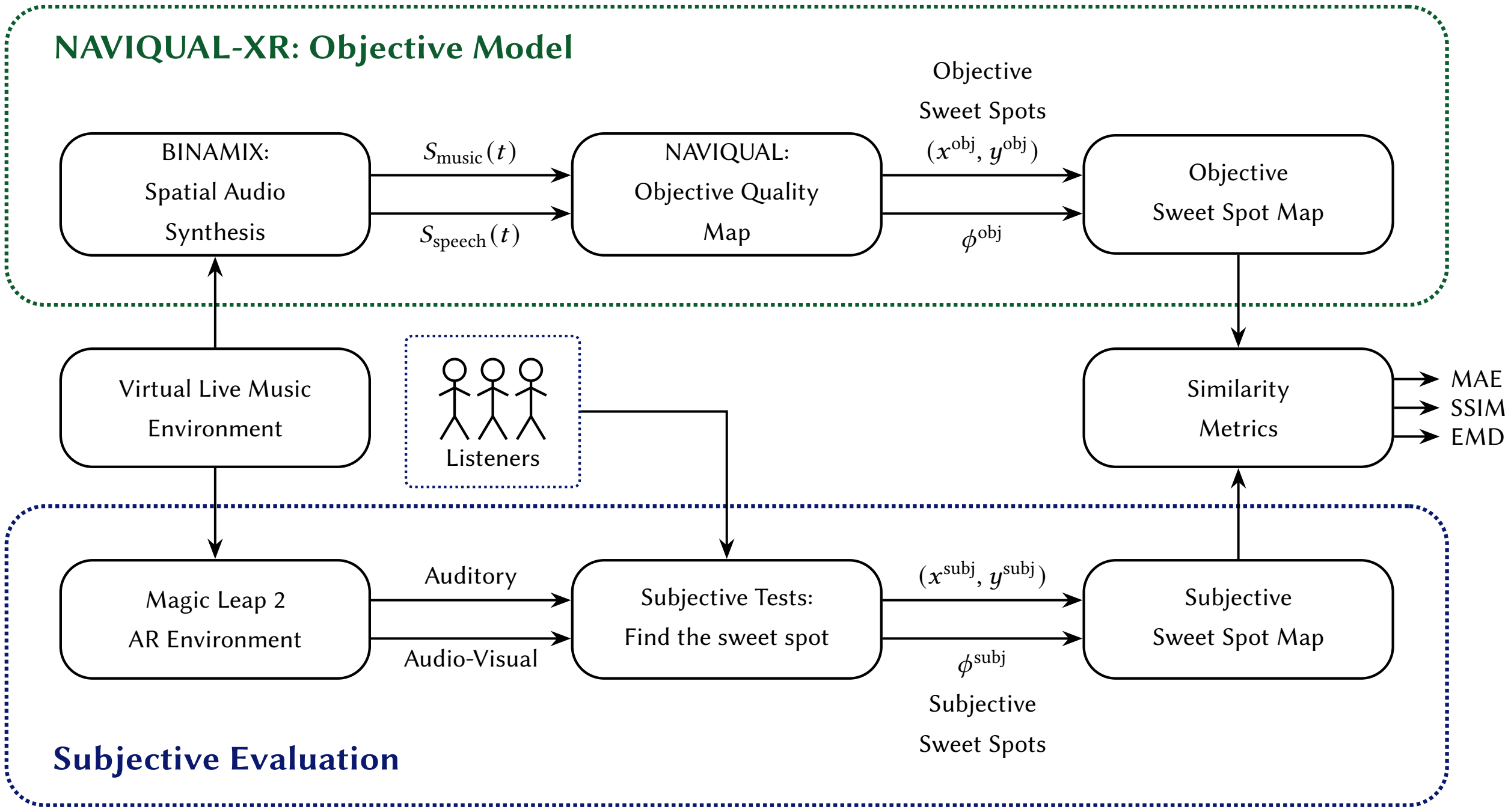
\begin{teaserfigure}
  \centering

\begin{tikzpicture}[
    >=Stealth,
    node distance=1.2cm and 2.2cm,
    block/.style={
        draw,
        rectangle,
        rounded corners=10pt,
        minimum width=3.4cm,
        minimum height=1.3cm,
        align=center,
        font=\sffamily\small,
        fill=white,
        line width=0.8pt
    },
    arr/.style={
        ->,
        line width=0.75pt
    },
    lbl/.style={
        font=\sffamily\small,
        inner sep=3pt
    }
]

    \node[block] (binamix) {BINAMIX:\\Spatial Audio\\Synthesis};
    \node[block, right=of binamix] (naviqual) {NAVIQUAL:\\Objective Quality\\Map};
    \node[block, right=of naviqual] (obj_map) {Objective\\Sweet Spot Map};

    \node[block, below=1.0cm of binamix] (vlme) {Virtual Live Music\\Environment};

    \node[block, below=1.0cm of vlme] (ml2) {Magic Leap 2\\AR Environment};
    \node[block, right=of ml2] (subj_tests) {Subjective Tests:\\Find the sweet spot};
    \node[block, right=of subj_tests] (subj_map) {Subjective\\Sweet Spot Map};

    \node[block] (sim_metrics) at (vlme -| obj_map) {Similarity\\Metrics};

    \coordinate (listener_center) at ($(vlme.east)!0.2!(subj_tests.west) + (0.9, 0.5cm)$);

    \begin{scope}[shift={(listener_center)}]
        \foreach \x in {-0.42, 0, 0.42} {
            \begin{scope}[xshift=\x cm, scale=0.85]
                \draw[line width=0.8pt] (0, 0.45) circle (0.14cm);
                \draw[line width=0.8pt] (0, 0.31) -- (0, -0.15);
                \draw[line width=0.8pt] (-0.2, 0.12) -- (0, 0.22) -- (0.2, 0.12);
                \draw[line width=0.8pt] (-0.18, -0.45) -- (0, -0.15) -- (0.18, -0.45);
            \end{scope}
        }
        \node[font=\sffamily\small] at (0, -0.58) (listener_text) {Listeners};
    \end{scope}

    \node[draw=subjblue, densely dotted, line width=0.8pt, rounded corners=2pt, 
          fit={(listener_center) ($(listener_center)+(0.85, 0.65)$) ($(listener_center)+(-0.85, -0.80)$)}] (listener_box) {};

    \draw[arr] ([yshift=6pt]binamix.east) -- node[above, lbl] {$S_{\mathrm{music}}(t)$} ([yshift=6pt]naviqual.west);
    \draw[arr] ([yshift=-6pt]binamix.east) -- node[below, lbl] {$S_{\mathrm{speech}}(t)$} ([yshift=-6pt]naviqual.west);

    \draw[arr] ([yshift=6pt]naviqual.east) -- node[above, lbl, align=center] {Objective \\ Sweet Spots\\$(x^{\mathrm{obj}}, y^{\mathrm{obj}})$} ([yshift=6pt]obj_map.west);
    \draw[arr] ([yshift=-6pt]naviqual.east) -- node[below, lbl] {$\phi^{\mathrm{obj}}$} ([yshift=-6pt]obj_map.west);

    \draw[arr] ([yshift=6pt]ml2.east) -- node[above, lbl] {Auditory} ([yshift=6pt]subj_tests.west);
    \draw[arr] ([yshift=-6pt]ml2.east) -- node[below, lbl] {Audio-Visual} ([yshift=-6pt]subj_tests.west);

    \draw[arr] ([yshift=6pt]subj_tests.east) -- node[above, lbl] {$(x^{\mathrm{subj}}, y^{\mathrm{subj}})$} ([yshift=6pt]subj_map.west);
    \draw[arr] ([yshift=-6pt]subj_tests.east) -- node[below, lbl, align=center] {$\phi^{\mathrm{subj}}$\\Subjective \\ Sweet Spots} ([yshift=-6pt]subj_map.west);

    \draw[arr] (vlme.north) -- (binamix.south);
    \draw[arr] (vlme.south) -- (ml2.north);
    \draw[arr] (obj_map.south) -- (sim_metrics.north);
    \draw[arr] (subj_map.north) -- (sim_metrics.south);

    \draw[arr] (listener_box.east) -| (subj_tests.north);

    \draw[arr] ([yshift=9pt]sim_metrics.east) -- ++(0.5,0) node[right, font=\sffamily\small] {MAE};
    \draw[arr] (sim_metrics.east) -- ++(0.5,0) node[right, font=\sffamily\small] {SSIM};
    \draw[arr] ([yshift=-9pt]sim_metrics.east) -- ++(0.5,0) node[right, font=\sffamily\small] {EMD};

    \begin{scope}[on background layer]
        \node[
            draw=objgreen,
            densely dotted,
            line width=1.2pt,
            rounded corners=12pt,
            fit={(binamix) (naviqual) (obj_map) ($(binamix.north)+(0,1.1cm)$) ($(binamix.south)+(0,-0.25cm)$) ($(binamix.west)+(-0.3cm,0)$) ($(obj_map.east)+(0.3cm,0)$)},
            inner sep=8pt
        ] (obj_container) {};
        
        \node[anchor=north west, font=\sffamily\bfseries\Large, text=objgreen, xshift=12pt, yshift=-6pt] 
            at (obj_container.north west) {NAVIQUAL-XR: Objective Model};

        \node[
            draw=subjblue,
            densely dotted,
            line width=1.2pt,
            rounded corners=12pt,
            fit={(ml2) (subj_tests) (subj_map) ($(ml2.north)+(0,0.3cm)$) ($(ml2.south)+(0,-1.1cm)$) ($(ml2.west)+(-0.3cm,0)$) ($(subj_map.east)+(0.3cm,0)$)},
            inner sep=8pt
        ] (subj_container) {};

        \node[anchor=south west, font=\sffamily\bfseries\Large, text=subjblue, xshift=12pt, yshift=6pt] 
            at (subj_container.south west) {Subjective Evaluation};
    \end{scope}

\end{tikzpicture}

  \caption{System overview of evaluating NAVIQUAL-XR. Listeners wearing Magic Leap 2 AR glasses were asked to find the sweet spot in AR live music environments. The resulting subjective sweet spot maps were then compared against the NAVIQUAL-XR objective maps using similarity metrics.}
  \Description{This figure depicts a system overview flowchart for evaluating NAVIQUAL-XR, split into a top green objective model box and a bottom blue subjective evaluation box. Both branches stem from a Virtual Live Music Environment block. The top branch processes BINAMIX synthesized audio through NAVIQUAL to generate objective sweet spot maps. The bottom branch uses a Magic Leap 2 AR environment to run auditory and audio-visual sweet spot search tests with human listeners. Both resulting sweet spot maps feed into a similarity metrics box outputting MAE, SSIM, and EMD scores.}
  \label{fig: naviqual-xr-graphic}
\end{teaserfigure}

\maketitle

\section{Introduction}
Augmented Reality (AR) enhances user interaction with physical surroundings by overlaying digital media---such as images, 3D objects, text, and acoustic signals---onto real-world environments. State-of-the-art AR systems are typically delivered via smart glasses \cite{KOUTROMANOS2023100028}, enabling seamless, real-time interaction between virtual and physical spaces. Users can interact with 3D elements via spatial gestures (e.g., touching buttons, pinching objects) while experiencing spatially synchronized sound fields \cite{Jot2021RenderingMetaverse}. While AR has seen broad adoption across commerce, education, and healthcare \cite{KOUTROMANOS2023100028}, its potential to transform live music consumption remains underexplored.


AR systems generate spatial audio by rendering acoustic sources onto binaural headphones worn by the user. By leveraging interaural difference cues, listeners can accurately localize virtual sound sources within their environment. When combined with head-tracking hardware, binaural audio dynamically adapts sound fields relative to head rotations \cite{Kailas2026}. AR extends these capabilities from stationary three-degrees-of-freedom head-tracked rendering (3-DoF: yaw, pitch, roll) to six-degrees-of-freedom spatial navigation (6-DoF: translation and rotation), yielding highly immersive and interactive acoustic environments.


Virtual live music environments can be synthesized in AR by placing multiple sound objects throughout a room \cite{Woodard2025}. In contemporary music production, audio tracks are commonly divided into discrete stems, typically categorized as \textit{vocals}, \textit{drums}, \textit{bass}, and \textit{other} instruments \cite{musdb18-hq}. Deep-learning-based source separation techniques make it possible to isolate these stems from a composite stereo mix \cite{Rouard2022HybridSeparation}. In AR applications, these individual stems serve as discrete spatial audio sources distributed across physical space, recreating a "live music" performance \cite{10.1145/3474349.3480218}. Furthermore, music producers have integrated AR into creative mixing workflows to manipulate audio stems directly in 3D space \cite{RiddershomBargum2023, Woodard2025}.


Despite the promising capabilities of AR, subjective evaluations of AR live music environments from the end-user perspective remains limited. Prior studies have primarily focused on perceptual evaluations under constrained, 3-DoF experimental setups \cite{Hirway2024, Grinn2024ChangesConcert, Guastamacchia2024}. Unlike standardized, static audio evaluation methodologies such as MUSHRA (MUlti Stimulus test with Hidden Reference and Anchor) \cite{itur_bs1534_1} or ACR (Absolute Category Rating) \cite{itut_p800_1_2016}, a universally accepted framework for assessing audio Quality of Experience (QoE) in interactive AR environments is currently lacking. Because AR systems generate dynamic audio that continuously adapts to user motion, designing controlled subjective evaluations poses unique challenges. Furthermore, AR environment design introduces numerous confounding variables, including the spatial configuration of sound sources, room acoustics, and room dimensions.


To address these virtual acoustic design challenges, the NAVIQUAL framework \cite{Tolentino2026naviqual} was proposed to estimate perceived QoE in virtual acoustic environments. NAVIQUAL computes objective quality maps without requiring physical deployment, and its predictions have been validated against subjective MUSHRA tests in blind listening conditions \cite{Tolentino2026naviqual}. However, the applicability and accuracy of NAVIQUAL have yet to be investigated and validated in 6-DoF navigable AR environments.


In traditional room acoustics, the ``sweet spot'' denotes the spatial position that delivers optimal listening quality. This is conventionally defined as the locus where sound waves from symmetric loudspeakers arrive at equal intensities and phase alignment, minimizing interaural discrepancies \cite{Merchel2009analysis}. Small translational displacements from this region introduce perceptible phase and level differences, degrading listening quality \cite{Conetta2014spatial}. Prior studies have focused on expanding this sweet spot by optimizing loudspeaker placement \cite{9287492} or by defining it as a bounded spatial region where perceived quality remains perceptually indistinguishable from a target reference \cite{9829924}.


In this work, we define the ``subjective sweet spot'' as the specific location and direction chosen by a listener that provides the optimal live music experience. This location is influenced by factors such as acoustic layout, navigation trajectory, environmental distractors, and individual stem mixing preferences. By aggregating data across multiple participants, we can establish a statistical ``average sweet spot'' analogous to Mean Opinion Scores (MOS) in standardized audio testing.


Our work pursues two main objectives: (1)~to establish how listeners navigate and select subjective sweet spots in AR live music environments; and (2)~to evaluate how effectively objective metrics predict these subjective sweet spots. Specifically, we formulate six research questions (RQs) divided across two core themes:


\noindent\textbf{Part I: Subjective Listener Experience}
\begin{enumerate}[label=\textbf{RQ\arabic*}:, leftmargin=*]
    \item How do listeners navigate toward their subjective sweet spots in auditory vs. audio-visual AR environments?
    \item To what extent do participants agree on a consistent subjective sweet spot location and direction?
    \item How do subjective sweet spot selections differ between purely auditory and audio-visual AR environments?
\end{enumerate}

\noindent\textbf{Part II: Objective Metric Evaluation}
\begin{enumerate}[label=\textbf{RQ\arabic*}:, resume, leftmargin=*]
    \item How can objective quality metrics be evaluated against human-selected subjective sweet spots?
    \item How accurately do objective metrics predict empirical subjective sweet spots?
    \item How robust are objective sweet spot predictions across variations in musical genre, distractor configuration, and visual cues?
\end{enumerate}

\section{Related Work}

\subsection{Potential of Augmented Reality}

Several recent works have leveraged extended reality (XR) technology, an umbrella term for virtual reality (VR) and AR technologies, to enhance human interaction with speech, audio, and music. In fact, a dedicated research framework called the \textit{Musical Metaverse} \cite{Turchet2023} has been proposed, highlighting the potential of XR for novel musical activities and inclusive experiences, among others. Researchers further proposed the promising avenues of XR to support the hard-of-hearing (HoH), combining context awareness \cite{Mehra2020}, listener intent prediction \cite{Mehra2020}, and acoustic environment matching \cite{Luberadzka2025} to improve their interactions with speech and music.

To transition these visions into practice, recent initiatives have focused on building data sets necessary for designing data-driven algorithms for egocentric research. Project Aria was an initiative by Meta that accelerated the creation of these data sets through the use of their glasses called \textit{Aria glasses} \cite{engel2023projectarianewtool}. Through the Aria glasses, egocentric multi-modal data sets have been produced to support algorithms for computer vision \cite{kong2025ariagen2pilot}, speech and audio \cite{donley2021easycomaugmentedrealitydataset}, and music research \cite{Ragano2025}. Notably, the EgoMusic dataset \cite{Ragano2025} captured multi-channel live music performances, demonstrating robust spatial audio capture across diverse theater locations. Similarly, the EasyCom dataset \cite{donley2021easycomaugmentedrealitydataset} targeted speech perception in noisy environments to mitigate the cocktail party problem. Together, these datasets highlight how multi-channel beamforming through wearable AR glasses improve audio quality and speech intelligibility \cite{Ragano2025, donley2021easycomaugmentedrealitydataset}. 

In parallel, developers have designed XR applications across accessibility, spatial experiences, and content creation. To improve sound accessibility for HoH individuals, visual and haptic substitutes have been deployed in VR \cite{Jain2021}, while spatialized AR systems like HoloBand \cite{Ivanyi2022} have been integrated into clinical speech pathology workflows. Spatial optimization algorithms such as Auptimize \cite{Cho2024Auptimize} have been designed to minimize localization errors, enabling more immersive AR experiences. For creators, applications in VR \cite{RiddershomBargum2023} and AR \cite{Woodard2025} have allowed music producers to mix stems in 6-DoF acoustic environments, addressing the spatial creativity constraints on traditional 2D interfaces.

While these data sets and applications demonstrate the utility of XR technology, there is still a need to conduct subjective studies with a broad set of participants to highlight the value and acceptance of the technology to end-users.

\subsection{Subjective Tests with Spatial Audio and AR}
Several recent studies have explored listener preferences on spatial audio and XR content. Existing subjective studies span a spectrum from traditional, blind listening tests to multi-modal and 6-DoF immersive assessments, highlighting key trade-offs between acoustic fidelity, visual integration, and cross-modal interactions.

Early perceptual studies primarily focused on purely auditory evaluations, isolating listener preferences without visual feedback. For example, head-related transfer function (HRTF) evaluations demonstrated that listeners preferred generalized mannequin HRTFs (e.g. KU100) over human HRTFs---and surprisingly, even their own personalized HRTF---in terms of overall perceptual quality \cite{Armstrong2018}. Similarly, investigations into loudspeaker configurations revealed that while scaling from stereo to 7.1 surround sound significantly increased perceived immersiveness, listeners still favored stereo formats for their familiarity and naturalness in music listening \cite{Riionheimo2024spatial1, Riionheimo2024spatial2}. However, by relying exclusively on blind listening scenarios, these studies omit the primary modality in AR: real-time visual context.

To address this multimodal gap, subsequent work introduced visual components using 360$^\circ$ videos and VR displays. Presenting 360$^\circ$ music content via VR headsets with spatial audio yielded higher ratings for presence, audio quality, and overall QoE compared to flat-screen stereo formats \cite{Holm2020}. Listener behavior metrics further reinforced these findings: integrating third-order ambisonics elevated visual attention and QoE, as measured by head pose tracking, pupil dilation, and heart rate dynamics \cite{Hirway2024}. Real-world physical spaces such as lecture halls have also been designed in VR, demonstrating that users can effectively envision themselves in these spaces for remote learning applications \cite{Guastamacchia2024}. Nevertheless, these multi-modal assessments were constrained to 3-DoF situations, where the listener's translational movement remains fixed.

More recent efforts have expanded subjective testing into 6-DoF environments and cross-modal interactions. In volumetric VR music experiences, integrating multi-sensory feedback, such as haptic audio-tactile gloves, significantly enhanced user experience across metrics like attractiveness, stimulation, and novelty \cite{Young2023}. On the algorithmic side, synthesized 6-DoF binaural room impulse responses (BRIRs) derived from single-location mono recordings have been assessed using listener discrimination tests \cite{arend2021six}. They found that while synthesis was promising, expert listeners can still distinguish between real and synthesized BRIRs \cite{arend2021six}.

In summary, we found that most subjective tests on spatial audio and AR were conducted through traditional blind listening tests and were mostly limited to 3-DoF scenarios. Recent subjective tests with 6-DoF environments relied on qualitative analysis through user experience questionnaires and lacked validation with objective metrics. Thus, there is a need to investigate the subjective experiences of listeners in 6-DoF environments, with a subjective test design that allows for objective analysis with quantitative data, and to evaluate how existing objective metrics align with these experiences. 

\subsection{Objective Metrics for Spatial Audio}
Objective metrics are designed to predict the subjective quality of an audio signal, reducing the cost of conducting multiple subjective tests on various conditions. Over the past decades, objective metrics have evolved from full-reference, signal-based standards to data-driven, non-intrusive models, and more recently, to spatial and binaural domains. 

Early standardized metrics focused primarily on single-channel speech quality. Full-reference algorithms such as Perceptual Evaluation of Speech Quality (PESQ) \cite{941023} and its successor Perceptual Objective Listening Quality Analysis (POLQA) \cite{POLQA} compare spectro-temporal representations of a degraded signal against a clean reference. To evaluate audio without access to reference signals, non-intrusive methods were introduced such as Deep Noise Suppression Mean Opinion Score (DNSMOS) \cite{9746108}, which utilized deep neural networks to independently predict speech quality, background noise level, and overall quality.

To improve generalization across diverse acoustic domains, subsequent objective metrics introduced advanced spectro-temporal features and self-supervised architectures. For instance, Virtual Speech Quality Objective Listener (VISQOL) \cite{Chinen2020} evaluates speech \cite{Hines2015} and audio \cite{Sloan2017VISQOLAudio} by measuring neurogram similarity (NSIM) between magnitude spectrograms, demonstrating high correlation with MUSHRA listening scores of expert listeners. For non-intrusive models, SCOREQ \cite{Ragano2024SCOREQ} adopts a contrastive regression strategy with triplet loss, learning robust distance representations that outperform state-of-the-art full-reference and no-reference metrics across diverse speech domains. 

As these aforementioned objective metrics do not account for spatial cues, specialized metrics have been developed for spatial and binaural domains. For instance, AMBIQUAL is a full-reference metric designed specifically for higher-order amnbisonic signals. Inspired from VISQOL, AMBIQUAL \cite{Narbutt2020} compares the phase spectrograms (phaseograms) of test and reference signals to predict the overall QoE and localization quality, strongly correlating with MUSHRA scores from expert listeners. This approach was subsequently adapted to binaural formats in BINAQUAL \cite{Shariat2026binaqual}, which similarly demonstrated high correlation with the subjective experience of expert listeners. More recently, NAVIQUAL \cite{Tolentino2026naviqual} deployed BINAQUAL on binaural signals synthesized within virtual acoustic environments, establishing strong correlations with MUSHRA scores in representative locations and directions in these environments.  

These existing objective metrics have yet to be thoroughly validated in dynamic, real-time 6-DoF AR settings with human movement. Thus, we leveraged AR technology in producing real-time 6-DoF binaural audio to conduct subjective tests of AR live music environments and to predict the QoE using existing objective metrics. We utilized the full 6-DoF suite in AR technologies, moving forward from traditional 3-DoF stationary evaluations. We also explored how the visual aspects influenced audio quality perception. We designed subjective studies that acquired more objective evaluations from the participants rather than relying on subjective questionnaires and discussions. Finally, we validated our experiments with both traditional and state-of-the-art objective metrics, offering useful insights on how we can predict the subjective experience of human listeners navigating in AR environments.

\section{Methodology}

A high-level overview of NAVIQUAL-XR is presented in Figure~\ref{fig: naviqual-xr-graphic}. We created 12 AR live music environments to be tested and evaluated by 25 participants. We collected their spatial coordinates over time for each scene, recording their final locations and directions as their subjective sweet spots. From these data, we produced subjective sweet spot maps that identified the most likely regions to be selected by a listener as the sweet spot. Finally, we evaluated and compared key objective metrics in their prediction capabilities of the sweet spots using similarity metrics against the subjective sweet spot maps.

\subsection{NAVIQUAL: Objective Quality Maps}

NAVIQUAL is a system designed to create an objective quality map of a virtual live music environment \cite{Tolentino2026naviqual}, as formally defined in Algorithm~\ref{alg: naviqual}. A virtual live music environment comprises three components: (1) music sources, (2) speech distractor sources, and (3) a designated listening area. Both music and speech distractor sources are characterized by their signal content and their spatial locations. The listening area defines the space over which the quality map is generated. Representative listener locations and directions are sampled within this area to synthesize binaural audio signals, which are then evaluated using objective quality metrics to produce the final quality map.

\begin{algorithm}[t]
\caption{NAVIQUAL: Objective Quality Map Generation}
\label{alg: naviqual}

\KwIn{%
  $\mathcal{M} = \{(C_m, L_m)\}$: Music sources (content, location)\;
  $\mathcal{D} = \{(C_d, L_d)\}$: Speech distractor sources (content, location)\;
  $\mathcal{P}$: Set of listener locations $(x, y)$\;
  $\Phi$: Set of listener directions $\phi$\;
  $\text{Metric}$: Selected objective quality metric ($\text{SNR}$ or $\text{LQ}$)\;
}
\KwOut{%
  $M_Q(p, \phi)$: Objective quality map over $(\mathcal{P}, \Phi)$ for the selected metric\;
}

\BlankLine
Initialize empty objective quality map $M_Q$\;

\ForEach{listener location $p \in \mathcal{P}$}{
    \ForEach{listener direction $\phi \in \Phi$}{
        \tcp{Synthesize binaural signals using Binamix}
        $S_{\text{music}}(t) \leftarrow \text{Binamix}(\mathcal{M}, p, \phi)$\;
        $S_{\text{speech}}(t) \leftarrow \text{Binamix}(\mathcal{D}, p, \phi)$\;
        
        \BlankLine
        \tcp{Compute and store selected objective quality metric}
        \eIf{$\text{Metric} = \text{SNR}$}{
            $M_Q(p, \phi) \leftarrow \text{LUFS}[S_{\text{music}}(t)] - \text{LUFS}[ S_{\text{speech}}(t)]$\;
        }{
            $M_Q(p, \phi) \leftarrow \text{BINAQUAL}[S_{\text{music}}(t) + S_{\text{speech}}(t), S_{\text{music}}(t)]$\;
        }
    }
}
\Return{$M_Q$}\;
\end{algorithm}




        
        

In \cite{Tolentino2026naviqual}, the virtual live music environments featured a $\qty{5}{\meter} \times \qty{5}{\meter}$ listening area, four frontal music sources (\textit{bass}, \textit{drums}, \textit{vocals}, and \textit{other}), and a single distractor source. Representative listener locations and directions were sampled from the environment at resolutions of \qty{0.5}{\meter} (translational) and $5^{\circ}$ (directional). Binaural audio signals were synthesized using Binamix \cite{Barry2025binamix}---a tool for synthesizing binaural audio from sound objects with azimuth and elevation parameters, which was extended in NAVIQUAL to support translation \cite{Tolentino2026naviqual}. Finally, listener QoE for these binaural signals was computed using two objective metrics: signal-to-noise ratio and localization quality.

The signal-to-noise ratio (SNR) is computed as the difference between the perceptual loudness---measured in Loudness Units Relative to Full Scale (LUFS)---of the music sources $S_{\text{music}}(t)$ and the distractor sources $S_{\text{speech}}(t)$. A high SNR indicates that the desired music dominates the distractors, generally yielding a higher music QoE. Meanwhile, localization quality (LQ) is computed using BINAQUAL \cite{Shariat2026binaqual}, a full-reference objective metric for binaural signals that measures phaseogram degradations relative to a lossless reference. In \cite{Tolentino2026naviqual}, the mixture of music and distractor sources $S_{\text{music}} (t) + S_{\text{speech}}(t)$ serves as the test signal, whereas the uninterrupted music $S_{\text{music}}(t)$ serves as the reference. LQ ranges from 0 to 1 and is directly proportional to and strongly correlated with perceived QoE, as demonstrated in \cite{Shariat2026binaqual, Tolentino2026naviqual}. Together, these two metrics evaluate binaural signals across two fundamental characteristics: intensity and phase.

\begin{figure}[t]
    \centering
    \subfloat[LQ contour map]{\includegraphics[width=0.4\textwidth]{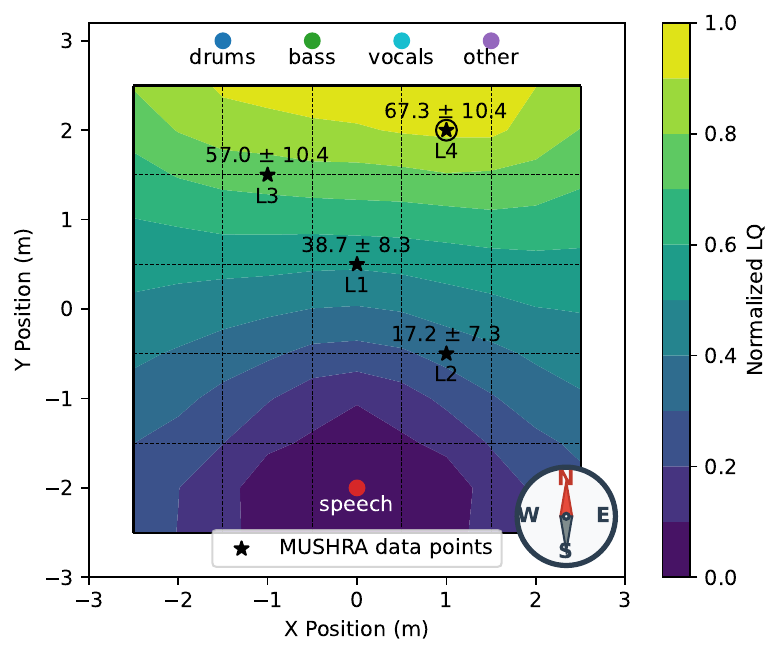}%
    \label{fig: naviqual-results-contour-1}}
    \hfil
    \subfloat[LQ contour map]{\includegraphics[width=0.4\textwidth]{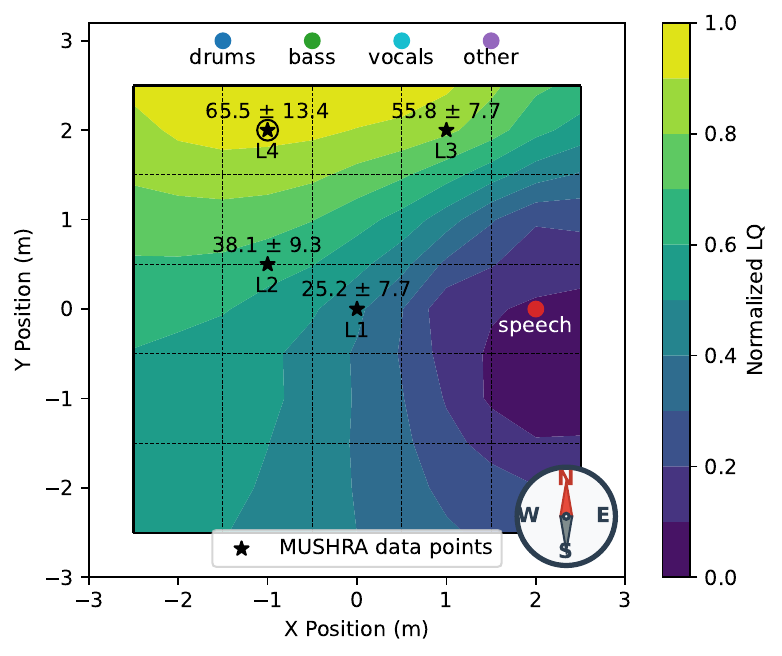}%
    \label{fig: naviqual-results-contour-2}}
    \hfil
    \subfloat[Polar plots]{\includegraphics[width=0.4\textwidth]{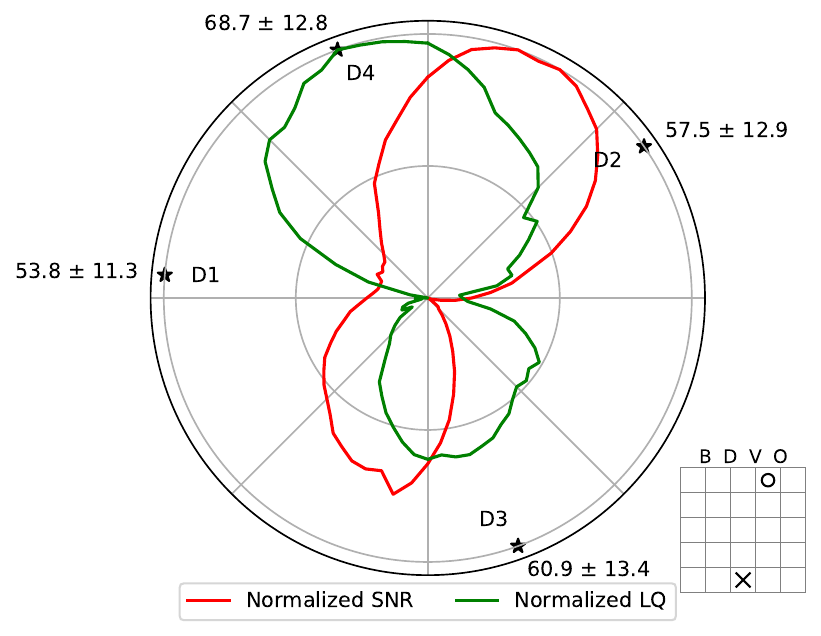}%
    \label{fig: naviqual-results-polar-1}}
    \hfil
    \subfloat[Polar plots]{\includegraphics[width=0.4\textwidth]{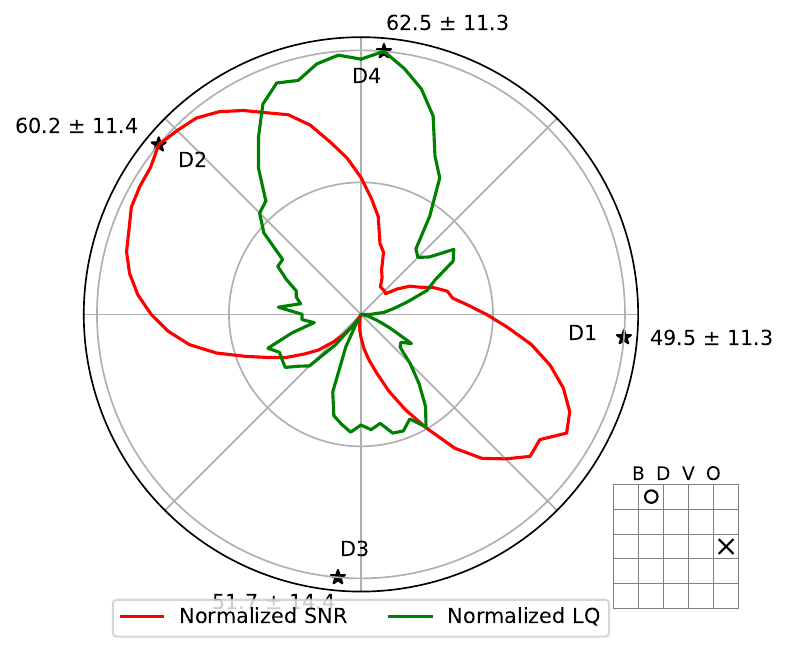}%
    \label{fig: naviqual-results-polar-2}}
    \hfil
    \caption{The objective quality contour maps (top row) and polar plots (bottom row) of a virtual live music environment. The bright-colored regions in the contour maps correspond to the locations with high subjective QoE. Meanwhile, the polar plots show the predicted optimal listening directions at a fixed location. The MUSHRA listening test results confirmed that the objective quality metrics strongly correlate with the subjective quality. Moreover, the LQ metric (green polar plot) surpassed the SNR metric (red polar plot) in predicting the optimal listening directions.}
    \Description{This figure displays four plots comparing objective metrics against subjective MUSHRA scores in a virtual live music environment. Top row subfigures (a) and (b) show 2D spatial contour maps of normalized LQ from 0.0 (blue, low) to 1.0 (yellow, high) with marked MUSHRA positions L1–L4. Bottom row subfigures (c) and (d) show 360-degree polar plots comparing Normalized SNR (red) and Normalized LQ (green) against directional subjective scores D1–D4.}
    \label{fig: naviqual-results}
\end{figure}

In NAVIQUAL \cite{Tolentino2026naviqual}, these two objective metrics were used to generate objective quality contour maps and polar plots as shown in Figure~\ref{fig: naviqual-results}, depicting the predicted subjective QoE at corresponding listener locations and directions within the environment. These maps were validated using representative listener locations and directions as test conditions in blind listening tests with 13 expert listeners \cite{Tolentino2026naviqual}, following a MUSHRA-style methodology \cite{itur_bs1534_1}. In Figure~\ref{fig: naviqual-results}, the representative locations and directions are marked by stars, with adjacent values showing average MUSHRA scores (on a 100-point scale) and 95\% confidence intervals. For the circled star, listener directions were varied at that fixed location to generate the polar plots in Figures~\ref{fig: naviqual-results-polar-1} and \ref{fig: naviqual-results-polar-2}. Overall, the LQ maps aligned closely with the MUSHRA scores, achieving an average Pearson correlation coefficient of $r = 0.93$ across 24 scenes encompassing diverse genres, distractor positions, and room acoustics \cite{Tolentino2026naviqual}. Furthermore, while both SNR and LQ predicted spatial QoE equally well, LQ significantly outperformed SNR in predicting optimal listening directions \cite{Tolentino2026naviqual}. For further details on the methodology and experimental findings, we refer the reader to \cite{Tolentino2026naviqual}.

Having established that NAVIQUAL objective quality maps strongly correlate with subjective ratings in static blind listening tests, we extend this framework in NAVIQUAL-XR to evaluate their predictive capabilities in augmented reality (AR) scenarios involving real-time, dynamic user navigation. NAVIQUAL streamlined the AR environment selection process for the subjective tests. We initially simulated several candidate AR environments and generated their corresponding objective quality maps. From these, we selected environments where the SNR and LQ maps exhibited distinct spatial profiles---spanning diverse acoustic content and scene setups---allowing us to compare the alignment of the objective metrics with the subjective experience. This eliminated the need to physically deploy and subjectively evaluate an exhaustive suite of AR candidate environments. 


\subsection{Design of the AR Environments}

Using Unity 6, we designed 12 AR scenes to be evaluated by the participants, as summarized in Table~\ref{tab: scene-setup}. In each scene, participants were free to navigate within a $\qty{5}{\meter} \times \qty{5}{\meter}$ bounded floor area marked by a $\qty{1}{\meter}$ grid. Four music sources---representing \textit{bass}, \textit{drums}, \textit{vocals}, and \textit{other}---were positioned at the front of the listening space at a height of $\qty{1.6}{\meter}$ and a distance of $\qty{0.5}{\meter}$ from the front edge. The music sources were set to a fixed intensity of $-30~\mathrm{LUFS}$ and remained in identical positions across all scenes.


\begin{table}[t]
    \caption{Scene setups for the subjective tests}
    \label{tab: scene-setup}
    \centering
    \begin{tabular}{
        >{\centering\arraybackslash}m{1.0cm}
        >{\centering\arraybackslash}m{2.6cm}
        >{\centering\arraybackslash}m{1.0cm}
        >{\centering\arraybackslash}m{2.6cm}
        >{\centering\arraybackslash}m{1.0cm}
        >{\centering\arraybackslash}m{2.6cm}
    }
    \toprule
    Scenes & Setup\tnote{1} & Scenes & Setup & Scenes & Setup \\
    \midrule
    \makecell[l]{S1-A \\ S1-AV} &    
    \includegraphics[width=2.5cm]{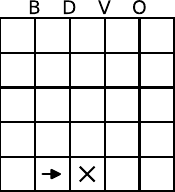} &
    \makecell[l]{S3-A \\ S3-AV} & 
    \includegraphics[width=2.5cm]{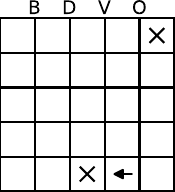} &
    \makecell[l]{S5-A \\ S5-AV} & 
    \includegraphics[width=2.5cm]{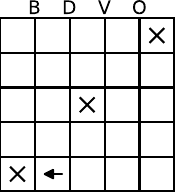} \\
    \midrule
    \makecell[l]{S2-A \\ S2-AV} & 
    \includegraphics[width=2.5cm]{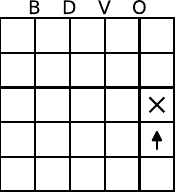}
    & \makecell[l]{S4-A \\ S4-AV} & 
    \includegraphics[width=2.5cm]{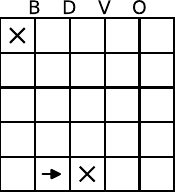}
    & \makecell[l]{S6-A \\ S6-AV} & 
    \includegraphics[width=2.5cm]{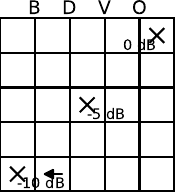} \\
    \bottomrule
    \end{tabular}
\end{table}

Between one and three speech distractor sources (marked as 'x' in Table~\ref{tab: scene-setup}) were positioned within the AR environments, with their locations varied across scenes. The first six scenes were auditory-only (\textit{S1-A} to \textit{S6-A}), where sound source positions were hidden from participants. Conversely, the remaining six scenes (\textit{S1-AV} to \textit{S6-AV}) were audio-visual counterparts identical to the first six, revealing the visual locations of the sound sources during navigation. Participants had no prior knowledge of the scene configurations until encountering these audio-visual scenes. The environments encompassed three song genres---\textit{Jazz} (\textit{S1}, \textit{S2}), \textit{Electronic} (\textit{S3}, \textit{S4}), and \textit{Pop Rock} (\textit{S5}, \textit{S6})---with an increasing number of distractors (one, two, and three, respectively). Distractor source intensities for scenes \textit{S1} through \textit{S5} were set to $0~\mathrm{dB}$ relative to the music sources. Scene \textit{S6} maintained the same distractor locations as \textit{S5} but varied their relative intensities to $0\text{ dB}$, $-5\text{ dB}$, and $-10\text{ dB}$. All source intensities attenuated with listener distance according to the inverse-clamped distance model \cite{OpenAL2005}. Music samples were sourced from the MUSDB18-HQ dataset \cite{musdb18-hq} and speech distractors from the clean subset of the MS-SNSD dataset \cite{AReddyAFramework}. A single song per genre and a set of speech samples were randomly selected and held constant across all participants.

\subsection{Subjective Test Design}
\begin{figure}[t]
    \centering
    \includegraphics[width=0.8\linewidth]{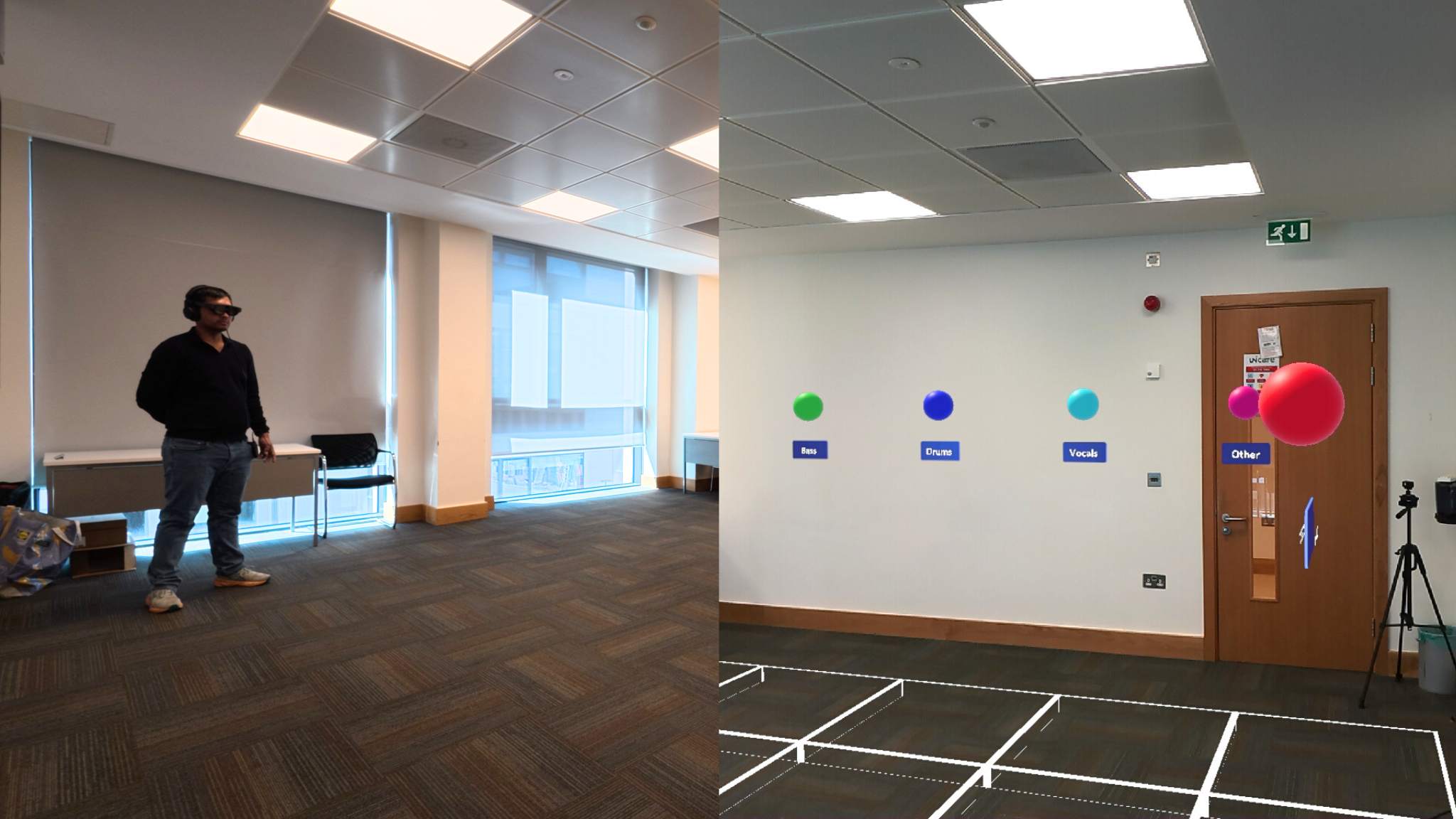}
    \caption{The third- and first-person view of a participant navigating an AR live music environment in a quiet room.}
    \Description{This figure displays a third- and first-person view of a person participating in the subjective tests. On the left, a person is standing in a room wearing a Magic Leap 2 AR glasses with headphones. On the right, the view from the glasses is shown with the rendered virtual live music environment consisting of the sound objects depicted as spheres.}
    \label{fig: room-setup}
\end{figure}

Figure~\ref{fig: room-setup} illustrates the physical setup for the subjective tests. In these tests, participants began navigating each AR environment from a standardized starting location and direction, indicated by the directional arrows in Table~\ref{tab: scene-setup}. Starting positions were deliberately selected in areas with low objective quality values, typically located near and facing a speech distractor source. Each starting position was marked on the floor by a green arrow. Participants were required to stand on this arrow and align with its direction before initiating the scene. From there, participants navigated toward the location and direction they perceived as offering the optimal music quality of experience---namely, the subjective ``sweet spot''.


To make the test more intuitive, participants were given two sequential tasks for each scene. First, they were instructed to find the location of the subjective ``sweet spot'' within the listening area. Once satisfied with their position, they remained in place and rotated to find the optimal listening direction. Upon confirming both location and direction, their chosen sweet spot was recorded, and they proceeded to the next scene until all twelve scenes were completed. To minimize order-presentation bias, we employed a randomized block design, randomizing scene order independently within the auditory and audio-visual blocks. Participant 6-DoF spatial tracking data were collected throughout navigation at a fixed sampling interval of $\qty{20}{\ms}$. Before the formal evaluation, participants completed a trial auditory scene to familiarize themselves with the environment and tasks. On average, participants took 18 minutes to complete the full test.


Subjective evaluations were conducted in a quiet room providing ample space for unconstrained navigation. Our custom AR software was deployed on Magic Leap 2 (ML2) AR glasses, which performed three primary functions: (1) augmenting the virtual live music environment onto the physical room, (2) tracking and logging 6-DoF spatial coordinates at each timestamp, and (3) rendering binaural audio in real time based on the participant's dynamic spatial coordinates.


For spatial audio rendering, we utilized the ML2-compatible Soundfield Audio plugin. Because this spatializer plugin is proprietary, we verified that its binaural signal outputs aligned with the spatializer used in NAVIQUAL \cite{Tolentino2026naviqual}. By recording a subset of binaural signals from the Soundfield Audio plugin and computing the SNR and LQ objective metrics on these outputs, we confirmed a strong correlation between the two systems ($R \ge 0.8779$ across various listener locations and directions; see Appendix A for details). This validated the direct applicability of the NAVIQUAL objective quality maps to our ML2-collected subjective data.


Audio was delivered through Audio-Technica ATH-M70x professional monitor headphones connected to the ML2 via a Bluetooth receiver with a $\qty{3.5}{\mm}$ output. All participants wore the same ML2 device and headphone model. To maintain a controlled experimental setup, we leveraged the Spatial Anchors feature of the ML2, ensuring that the physical location and orientation of the virtual environment remained persistent across all sessions.



We recruited 25 participants (20 male, 5 female) representing various age brackets: $18-25~\mathrm{years}$ ($n = 6$), $26-35~\mathrm{years}$ ($n=14$), $36-45~\mathrm{years}$ ($n=3$), and $46-55~\mathrm{years}$ ($n=2$). All participants confirmed that they have normal hearing without a hearing impairment diagnosis. Background questionnaires indicated that 21 participants had prior AR/VR experience, 13 had participated in subjective listening tests, and 9 had formal training in audio engineering, music, or acoustics---yielding a balanced cohort of naive and expert listeners. 

\subsection{Subjective Sweet Spot Maps}

Following the subjective tests, we generated a subjective sweet spot map to identify the most probable sweet spot locations and directions in our AR environments. These maps enabled us to evaluate the agreement between the participants' subjective choices and the objective quality maps produced by NAVIQUAL \cite{Tolentino2026naviqual}. First, we extracted the tracking data logged by the ML2 glasses, comprising the listener's location and direction at every timestamp for each participant. The subjective sweet spot $\mathbf{x}_i = [x_i, y_i, \phi_i]^\mathrm{T}$ for the $i$-th participant was computed as the median of their final 50 sample points (1 second of navigation), thereby mitigating recording jitter. From these individual sweet spots, we constructed a continuous subjective sweet spot map $\text{SQ}_{\mathrm{ss}}(\mathbf{x}) \in [0, 1]$ as a function of listener location $(x, y)$ and direction $\phi$. Higher $\text{SQ}_{\mathrm{ss}}(\mathbf{x})$ values indicate a higher likelihood of listeners positioning and orienting themselves at that point. Representative subjective sweet spot maps are illustrated in Figures~\ref{fig: sweetspotmap-subj-s1-av} and \ref{fig: sweetspotmap-subj-s4-a}.


We computed $\text{SQ}_{\mathrm{ss}}(\mathbf{x})$ using the Mahalanobis distance across all points $(x, y, \phi)$ in the environment. The Mahalanobis distance was selected because it measures how close an arbitrary point is to the underlying distribution of subjective sweet spots obtained from our tests while explicitly accounting for the variance and covariance of the user data. Because the listener direction $\phi$ is angular and non-linear, we decomposed it into continuous components $\cos(\phi)$ and $\sin(\phi)$, yielding a 4D feature vector $\mathbf{x} = [x, y, \cos(\phi), \sin(\phi)]^\mathrm{T}$ for every point. Given the subjective sweet spots $\mathbf{x}_i = [x_i, y_i, \cos(\phi_i), \sin(\phi_i)]^\mathrm{T}$ for $N$ participants, the Mahalanobis distance $D_M(\mathbf{x})$ at any point $\mathbf{x}$ is defined as:


\begin{equation}
    D_M(\mathbf{x}) = \sqrt{(\mathbf{x} - \bm{\mu})^T\Sigma^{-1}(\mathbf{x} - \bm{\mu})}
    \label{eqn: mahalanobis-distances}
\end{equation}


where the sample mean vector $\boldsymbol{\mu}$ and sample covariance matrix $\boldsymbol{\Sigma}$ are given by:

\begin{align}
    \bm{\mu} &= \frac{1}{N} \sum_{i=1}^N \mathbf{x}_i & \quad
    \Sigma &= \frac{1}{N-1} \sum_{i=1}^N (\mathbf{x}_i - \bm{\mu})(\mathbf{x}_i - \bm{\mu})^T
    \label{eqn: mean-covariance}
\end{align}

The subjective sweet spot map $\text{SQ}_{\mathrm{ss}}(\mathbf{x}) \in [0, 1]$ was evaluated by applying a Gaussian decay kernel to the Mahalanobis distances followed by min-max normalization:

\begin{equation}\text{SQ}_{\mathrm{ss}}(\mathbf{x}) = \frac{f(\mathbf{x}) - \min f}{\max f - \min f}, \quad \text{where } f(\mathbf{x}) = \exp\left(-\frac{1}{2} D_M^2(\mathbf{x})\right)
\label{eqn: mahalanobis-scores}
\end{equation}

\subsection{Correlating Subjective and Objective Quality}
\label{sec: methods-correlating}

Once the subjective sweet spot maps were constructed from the test results, we evaluated their alignment with the objective quality maps using similarity metrics. These metrics enabled us to determine which objective metric aligns more closely with user experience. Our objective quality maps were generated over a grid of $11 \times 11 \times 72 = 8,712$ evaluation points, with a spatial resolution of $0.5\text{ m}$ ($x$ and $y$ coordinates) and an angular resolution of $5^\circ$ ($\phi$). Accordingly, we sampled the identical set of $8,712$ points from the subjective sweet spot maps as input for similarity calculations.


To ensure both map types resulted from a unified processing pipeline, we applied an analogous transformation to the objective quality maps by computing Mahalanobis distances relative to a subset of objective points, as formally defined in Algorithm~\ref{alg: naviqual_xr}. Specifically, objective sweet spots were extracted from the quality maps (as in Figure~\ref{fig: naviqual-results}) by applying a threshold $T$ through a two-step process mirroring the subjective test methodology:

\begin{enumerate}[leftmargin=*, itemsep=2pt, topsep=2pt]
    \item \textbf{Location Selection:} Objective sweet spot locations $(x_i^\mathrm{obj}, y_i^\mathrm{obj})$ were selected where the normalized quality score facing forward ($\phi = 0^\circ$) exceeded $T$.
    \item \textbf{Direction Selection:} At each selected location $(x_i^\mathrm{obj}, y_i^\mathrm{obj})$, quality scores across all orientations were re-normalized, and directions $\phi_i^\mathrm{obj}$ exceeding $T$ were retained.
\end{enumerate}

\begin{algorithm}[t]
\caption{Objective Sweet Spot Map Generation}
\label{alg: naviqual_xr}

\KwIn{%
  $M_Q(p, \phi)$: Objective quality map from NAVIQUAL\;
  $T$: Sweet spot selection threshold\;
}
\KwOut{%
  $Q_{\text{ss}}(p, \phi)$: Normalized objective sweet spot map over $(\mathcal{P}, \Phi)$\;
}

\BlankLine
Extract location set $\mathcal{P}$ and direction set $\Phi$ from $M_Q(p, \phi)$\;

\BlankLine
\tcp{Step 1: Obtain objective sweet spots via two-step thresholding}
Initialize empty set of sweet spot vectors $\mathcal{S}_{\text{obj}} \leftarrow \emptyset$\;
\ForEach{location $p = (x, y) \in \mathcal{P}$}{
    \If{$M_Q(p, \phi=0) \ge T$}{
        $Q_{\text{dir}}(\phi) \leftarrow \text{Re-normalize } M_Q(p, \phi) \text{ across } \phi \in \Phi \text{ to } [0, 1]$\;
        \ForEach{direction $\phi \in \Phi$}{
            \If{$Q_{\text{dir}}(\phi) \ge T$}{
                $\mathcal{S}_{\text{obj}} \leftarrow \mathcal{S}_{\text{obj}} \cup \{[x, y, \cos(\phi), \sin(\phi)]^\mathrm{T}\}$\;
            }
        }
    }
}

\BlankLine
\tcp{Step 2: Compute Mahalanobis distances of the environment}
Compute mean vector $\boldsymbol{\mu}_{\text{obj}}$ and covariance matrix $\boldsymbol{\Sigma}_{\text{obj}}$ from $\mathcal{S}_{\text{obj}}$\;
\ForEach{location $p = (x, y) \in \mathcal{P}$ and direction $\phi \in \Phi$}{
    $\mathbf{x} \leftarrow [x, y, \cos(\phi), \sin(\phi)]^\mathrm{T}$\;
    $D_M(\mathbf{x}) \leftarrow \sqrt{(\mathbf{x} - \boldsymbol{\mu}_{\text{obj}})^\mathrm{T} \boldsymbol{\Sigma}_{\text{obj}}^{-1} (\mathbf{x} - \boldsymbol{\mu}_{\text{obj}})}$\;
    $Q_{\text{ss}}(p, \phi) \leftarrow \exp\left(-\frac{1}{2} D_M^2(\mathbf{x})\right)$\;
}

\BlankLine
\tcp{Step 3: Normalize map values to $[0, 1]$}
Min-max normalize $Q_{\text{ss}}(p, \phi)$ over all $p \in \mathcal{P}$ and $\phi \in \Phi$\;

\Return{$Q_{\text{ss}}$}\;
\end{algorithm}

\begin{figure}[t]
    \centering
    \subfloat[SNR quality map]{\includegraphics[width=0.4\textwidth]{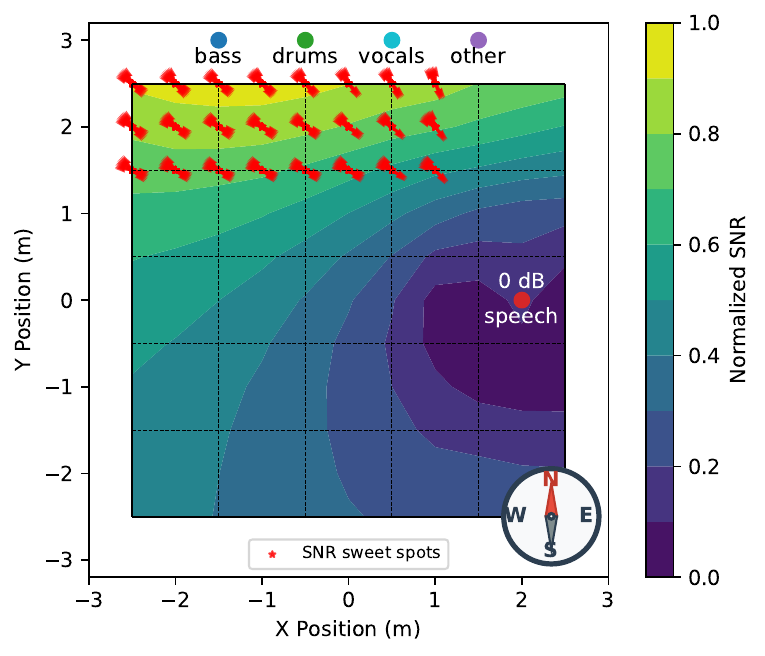}
    \label{fig: snr-quality}}
    \hfil
    \subfloat[LQ quality map]{\includegraphics[width=0.4\textwidth]{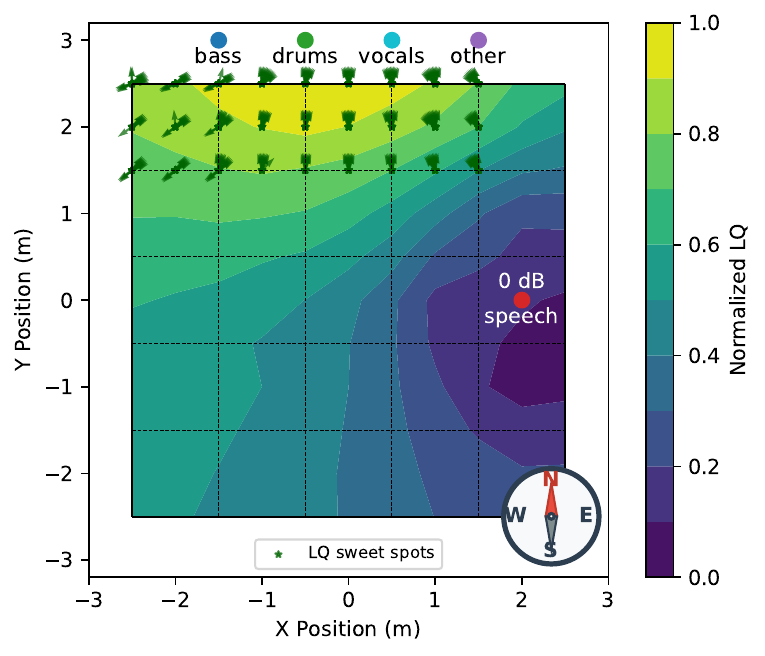}
    \label{fig: lq-quality}}
    \hfil
    \subfloat[SNR sweet spot map]{\includegraphics[width=0.4\textwidth]{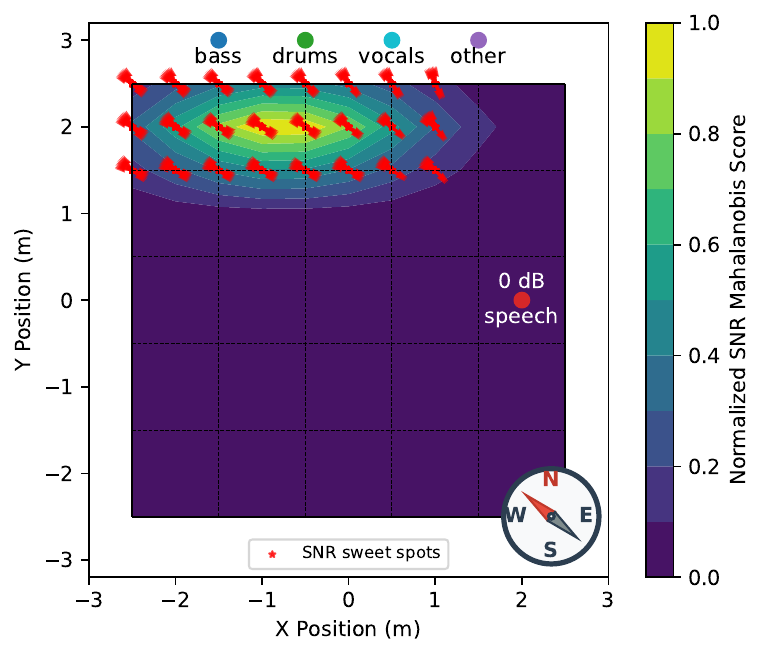}
    \label{fig: snr-ss}}%
    \hfil
    \subfloat[LQ sweet spot map]{\includegraphics[width=0.4\textwidth]{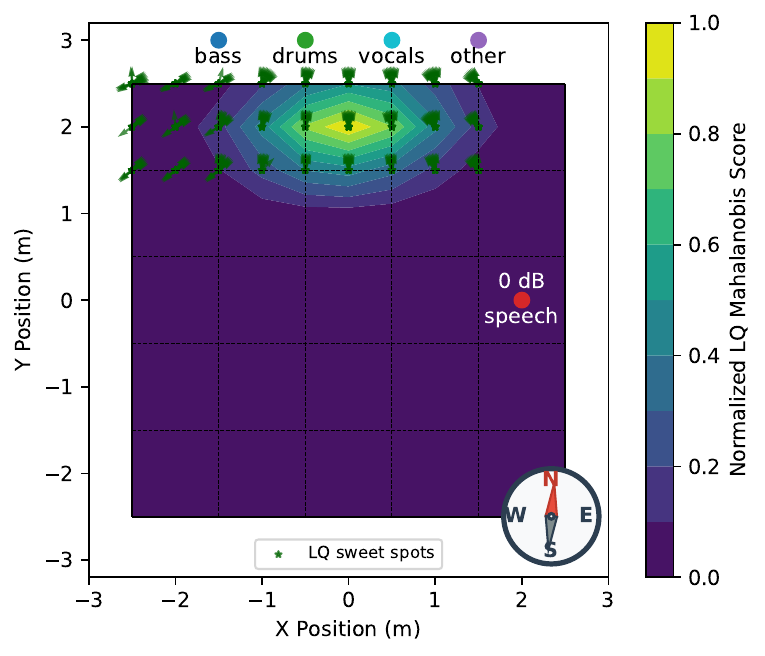}
    \label{fig: lq-ss}}%

    \hfil
    \caption{The objective quality maps (top row) and the resulting sweet spot maps (bottom row) based on the objective sweet spot data points. The compass arrows in the sweet spot maps correspond to the direction containing the highest sweet spot map value. We can see that the bright-colored regions in the sweet spot maps correspond to the center of the objective sweet spot locations, while the compass directions correspond to the concensus objective sweet spot directions.}
    \Description{This figure presents four 2D spatial maps in a 2x2 grid comparing objective audio metrics and derived sweet spot locations from X=-3m to 3m and Y=-3m to 3m. Four target audio sources (bass, drums, vocals, other) sit at the top boundary, and a 0 dB speech distractor sits at the right at X=2m, Y=0m. The top row shows broad quality contour gradients for Normalized SNR (a) and Normalized LQ (b). The bottom row shows localized Normalized Mahalanobis Score sweet spot maps for SNR (c) and LQ (d), featuring concentrated bright peaks around X=0m, Y=2m.}
    \label{fig: quality-sweetspot-maps}
\end{figure}

Figure \ref{fig: quality-sweetspot-maps} compares the raw objective quality maps with the resulting objective sweet spot maps derived from this procedure. In the objective sweet spot maps (Figures~\ref{fig: snr-ss} and \ref{fig: lq-ss}), high values are concentrated primarily within the sweet spot region, while remaining areas approach zero. Conversely, the raw objective quality maps span a continuous range from 0 to 1 across the entire space, capturing gradations from "bad" to "excellent" quality. We prioritized the objective sweet spot maps for similarity comparisons because participants were explicitly tasked with finding an optimal (``excellent'') spot; however, non-selected regions were implicitly evaluated as sub-optimal (``not excellent'') during navigation.


We evaluated map alignment using three similarity metrics:

\begin{enumerate}[leftmargin=*, itemsep=2pt, topsep=2pt]
    \item \textbf{Mean Absolute Error (MAE):} Measures pairwise point-to-point absolute differences between normalized $[0, 1]$ subjective and objective maps. Lower MAE indicates higher predictive accuracy. However, because MAE relies on point-by-point comparison, minor spatial shifts heavily penalize the score.
    \item \textbf{Structural Similarity Index Measure (SSIM):} Evaluates structural and gradient similarity between maps within a sliding window, rendering it robust to minor spatial displacements. SSIM ranges from 0 to 1, where higher values denote greater structural alignment.
    \item \textbf{Earth Mover's Distance (EMD):} Measures the minimum "work" required to transform one spatial distribution into another. EMD explicitly accounts for physical distance and positional offset between sweet spot regions, with lower values indicating higher similarity.
\end{enumerate}


\section{Results and Discussion}

We present our experimental results structured around the research questions outlined previously. We also provide the entire set of figures and animations for all evaluated scenes in our GitHub repository\footnote{https://github.com/QxLabIreland/naviqual-xr}, as only a representative subset of static figures is presented in this paper.

\subsection{Navigation Behavior}
\label{sec: results-navigation}
This section addresses RQ1: How do listeners navigate toward their subjective sweet spots in auditory vs audio-visual AR environments? We analyzed the participants' navigation trajectories and compared them across different scene setups in auditory and audio-visual conditions. Then, we conducted statistical analysis to determine how the environmental conditions affected their path efficiency towards their sweet spots.  

\begin{figure*}[t]
    \centering
    \subfloat[S1-A trajectories]{\includegraphics[width=0.33\textwidth]{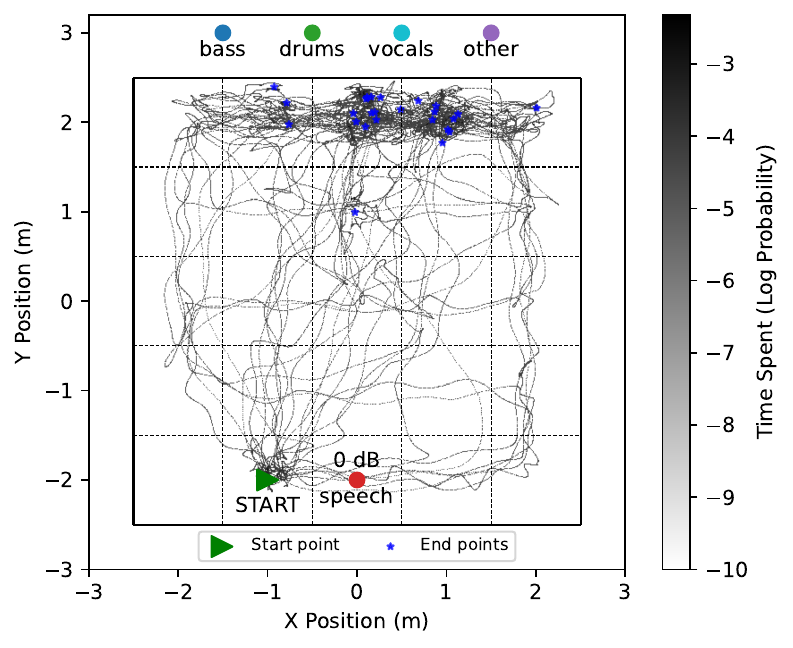}%
    \label{fig: grid-distribution-s1-a}}
    \subfloat[S3-A trajectories]{\includegraphics[width=0.33\textwidth]{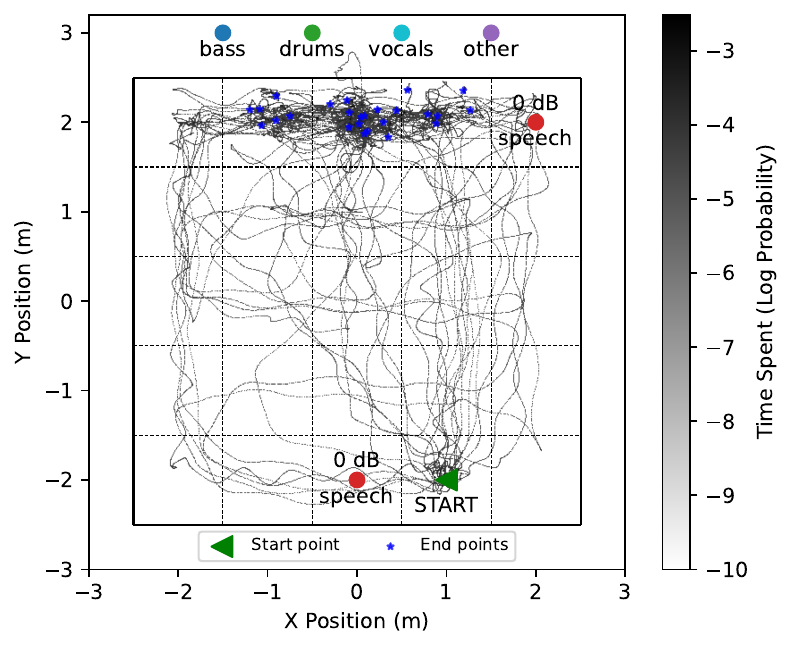}%
    \label{fig: grid-distribution-s3-a}}
    \subfloat[S5-AV trajectories]{\includegraphics[width=0.33\textwidth]{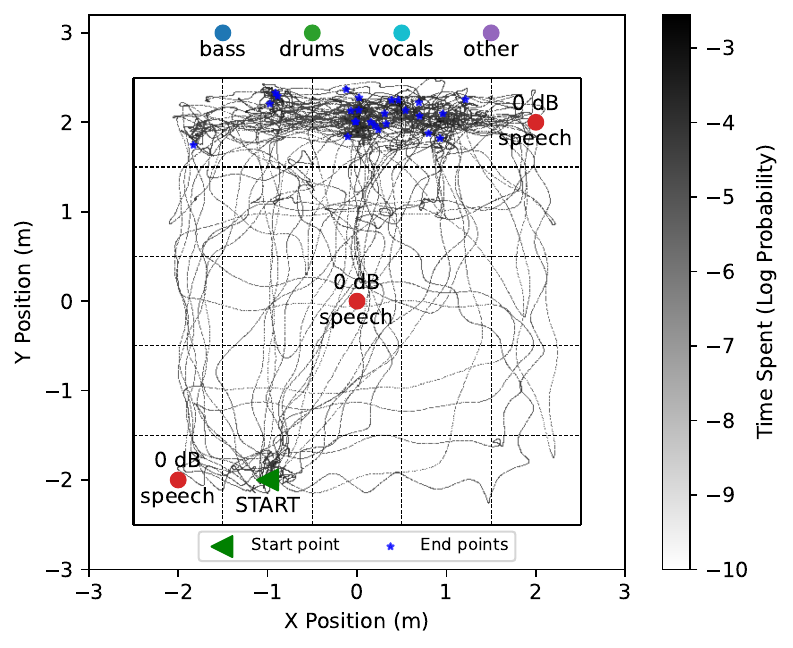}%
    \label{fig: grid-distribution-s5-a}}
    \hfil
    \break
    \subfloat[S5-A trajectories]{\includegraphics[width=0.33\textwidth]{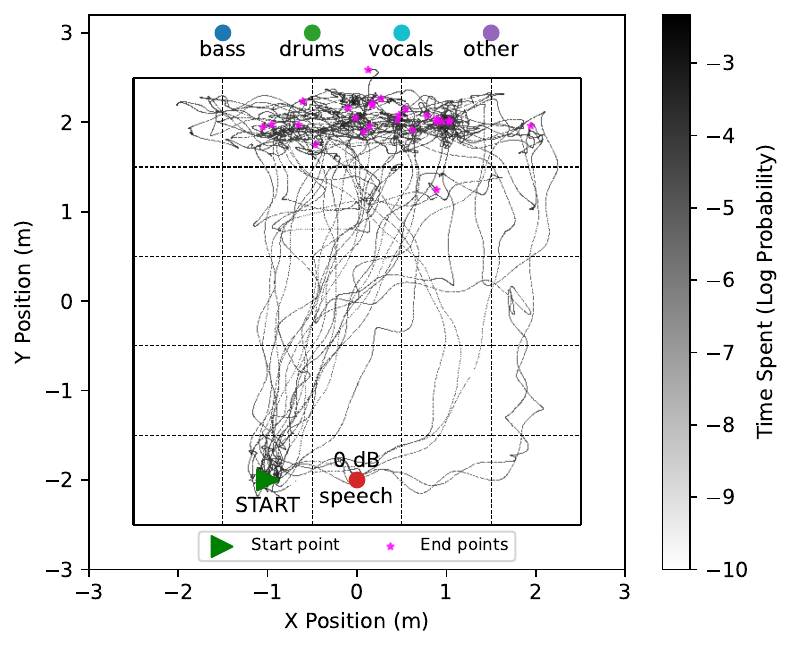}%
    \label{fig: grid-distribution-s1-av}}
    \subfloat[S1-A trajectories]{\includegraphics[width=0.33\textwidth]{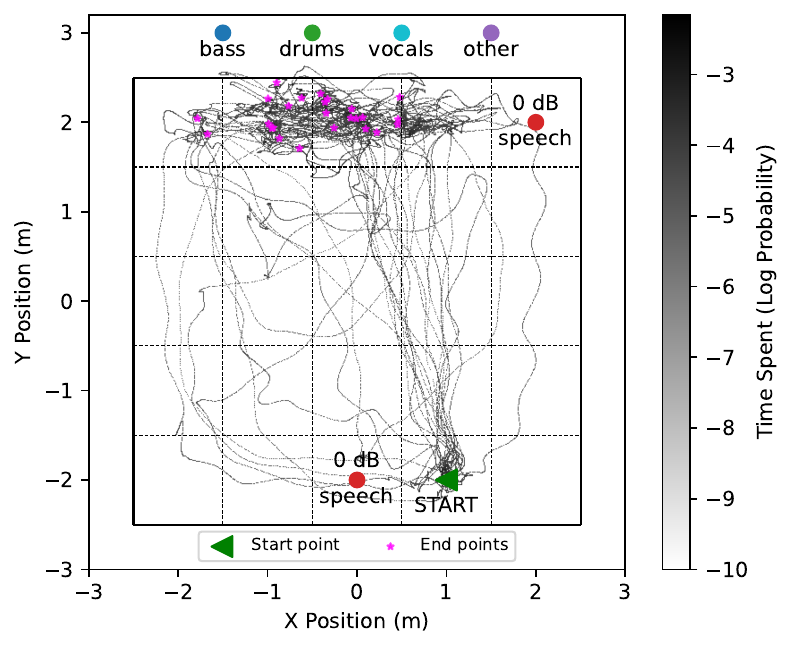}%
    \label{fig: grid-distribution-s3-av}}
    \subfloat[S5-AV trajectories]{\includegraphics[width=0.33\textwidth]{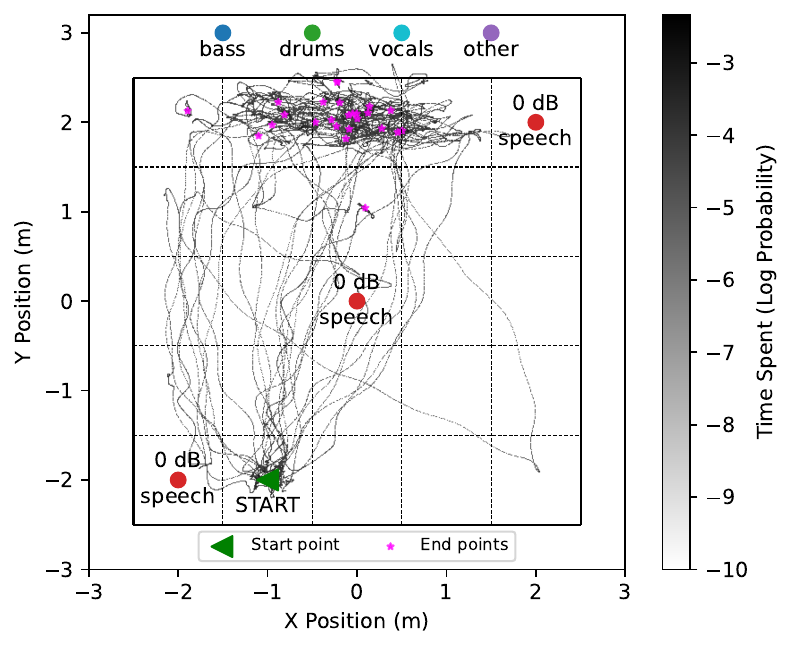}%
    \label{fig: grid-distribution-s5-av}}
    \caption{Participant navigation trajectories from the initial position to selected subjective sweet spots. The top (a--c) and bottom (d--f) rows depict auditory and audio-visual environments, respectively, across an increasing number of distractor sources. Audio-visual conditions yielded direct navigation paths, whereas auditory conditions prompted broader spatial exploration.}
    \Description{This figure shows six 2D spatial trajectory plots in a 2x3 grid comparing participant movement paths from start locations to final sweet spots in auditory (top row a–c) versus audio-visual environments (bottom row d–f) across 1, 2, and 3 speech distractors. Target music sources (bass, drums, vocals, other) sit at Y=3m. Each plot includes a grayscale colorbar for log time spent from -3 (dark, high) to -10 (light, low). Auditory conditions show diffuse, wide exploratory loops across the 5m x 5m area, whereas audio-visual conditions yield highly direct, linear paths toward the final spots.}
    \label{fig: grid-distribution}
\end{figure*}

Figure~\ref{fig: grid-distribution} illustrates participant trajectories from the origin to their selected subjective sweet spots. Trajectories are visualized using grayscale lines, where line intensity corresponds to participant residence time aggregated over a $0.1\,\text{m} \times 0.1\,\text{m}$ spatial grid. All participants initiated navigation from an identical starting position and orientation, indicated by the green arrow. The top row (Figures~\ref{fig: grid-distribution-s1-a}, \ref{fig: grid-distribution-s3-a}, and \ref{fig: grid-distribution-s5-a}) presents trajectories in auditory environments, whereas the bottom row (Figures~\ref{fig: grid-distribution-s1-av}, \ref{fig: grid-distribution-s3-av}, and \ref{fig: grid-distribution-s5-av}) depicts corresponding audio-visual conditions.


Auditory environments produced more stochastic trajectories, indicating that audio cues alone prompted greater spatial exploration. Conversely, audio-visual environments yielded more direct, linear trajectories, suggesting that visual cues enabled participants to identify their sweet spots more efficiently. Navigation patterns also varied according to the number and spatial arrangement of distractor sources. In Figures~\ref{fig: grid-distribution-s5-av} and \ref{fig: grid-distribution-s3-av}, participants consistently avoided the right portion of the listening space due to the presence of speech distractors. In contrast, in Figure~\ref{fig: grid-distribution-s1-av}, trajectories bypassed lateral regions and converged directly along the forward central axis toward the primary sweet spot.

\begin{figure}[t]
\centering
\includegraphics[width=0.75\textwidth]{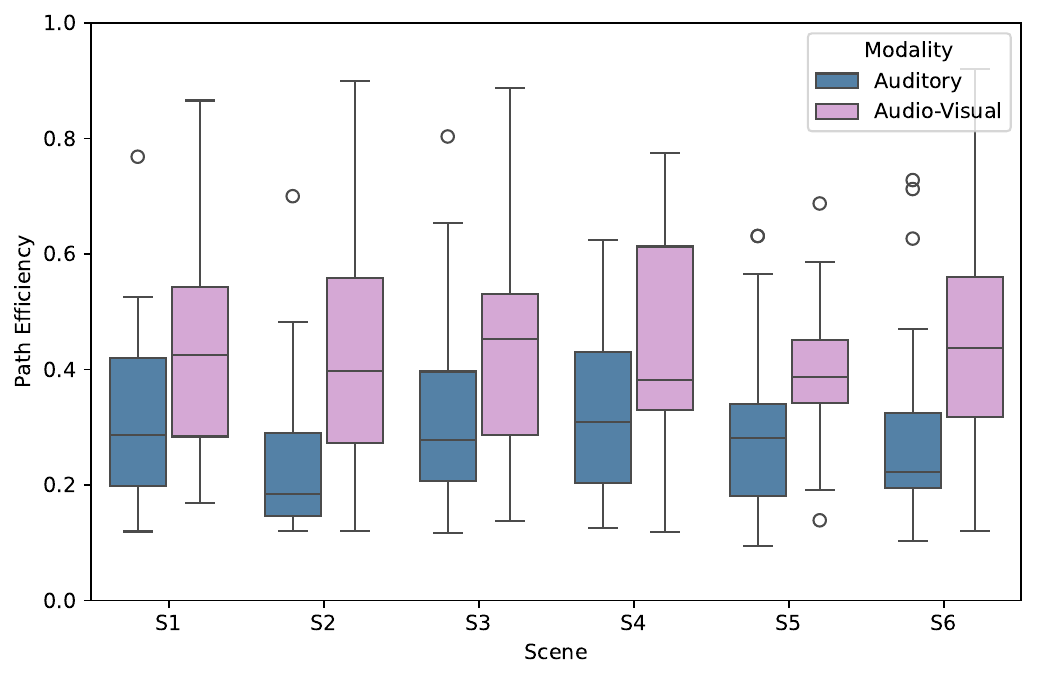}
\caption{Trajectory path efficiency---defined as the ratio of shortest path distance to total distance traversed---aggregated across participants for all scenes (S1--S6) and modalities. Audio-visual conditions (pink) yielded significantly higher path efficiency than auditory conditions (blue) ($p = 1.5 \times 10^{-5}$). Conversely, varying scene configurations produced no statistically significant change in path efficiency ($p = 0.117$).}
\Description{This figure shows a grouped box plot displaying path efficiency from 0.0 to 1.0 across six scenes (S1–S6) for Auditory (blue) and Audio-Visual (pink) modalities. Across all scenes, the Audio-Visual condition achieves higher path efficiency medians than the Auditory-only condition ($p = 1.5 \times 10^{-5}$), while scene variation shows no statistically significant effect ($p = 0.117$).}
\label{fig: path-efficiency}
\end{figure}

To quantify navigation directness, we evaluated trajectory path efficiency---defined as the ratio of the shortest path distance to the total distance traversed (where values closer to $1.0$ indicate more direct movement). Figure~\ref{fig: path-efficiency} presents box plots of path efficiency across all scenes and modalities. A two-way repeated-measures analysis of variance (RM-ANOVA) revealed a statistically significant main effect of cue modality on path efficiency ($p = 3.0 \times 10^{-5}$). Post-hoc Tukey’s HSD tests confirmed that audio-visual environments yielded significantly higher path efficiency than auditory conditions ($p = 1.5 \times 10^{-5}$), confirming that visual cues enabled participants to locate sweet spots more directly. In contrast, acoustic scene variation did not significantly affect path efficiency ($p = 0.117$), indicating that participants encountered comparable navigation challenges regardless of sound scene complexity.


\subsection{Participant Sweet Spots}
\label{sec: results-sweetspots}
This section addresses two research questions: To what extent do participants agree on a consistent subjective sweet spot location and direction (RQ2), and how do subjective sweet spot selections differ between purely auditory and audio-visual AR environments (RQ3)? We aggregated the participants' selected sweet spots to identify the concensus location and direction per scene and modality, and analyzed how these concensus spots varied according to environmental conditions.

\begin{figure*}[t]
    \centering
    \subfloat[S2-AV sweet spots]{\includegraphics[width=0.33\textwidth]{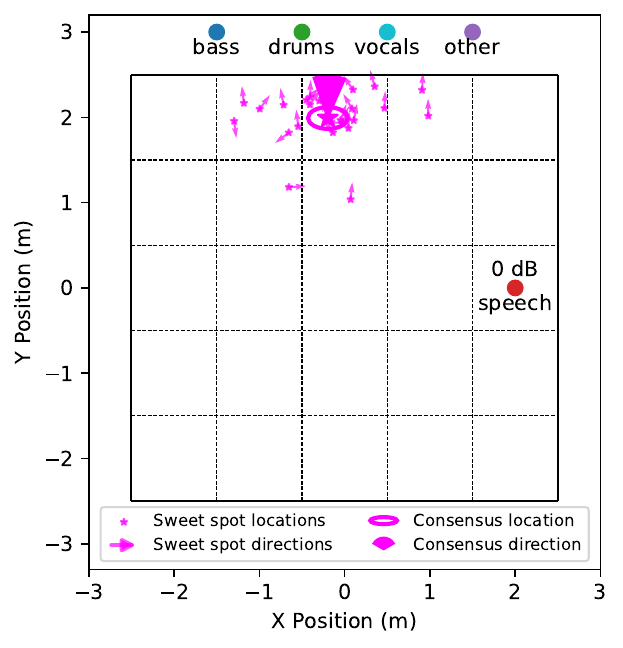}%
    \label{fig: sweetspots-s2-av}}
    \hfil
    \subfloat[S4-AV  sweet spots]{\includegraphics[width=0.33\textwidth]{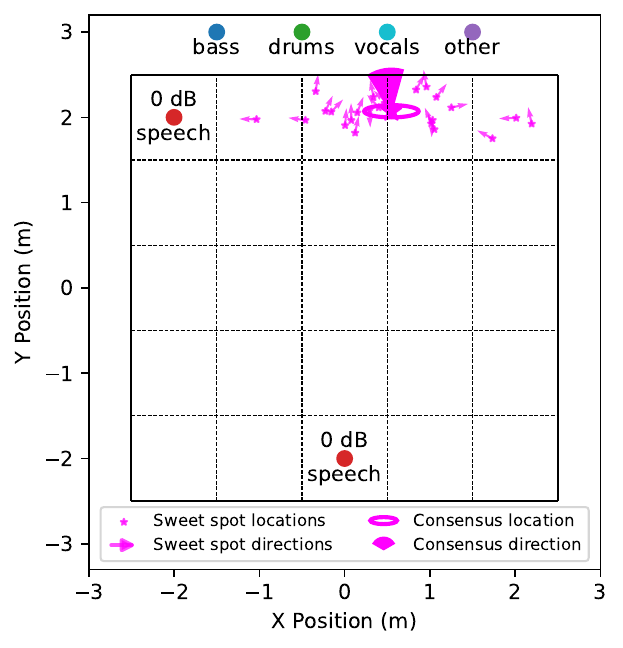}%
    \label{fig: sweetspots-s4-av}}
    \subfloat[S6-AV sweet spots]{\includegraphics[width=0.33\textwidth]{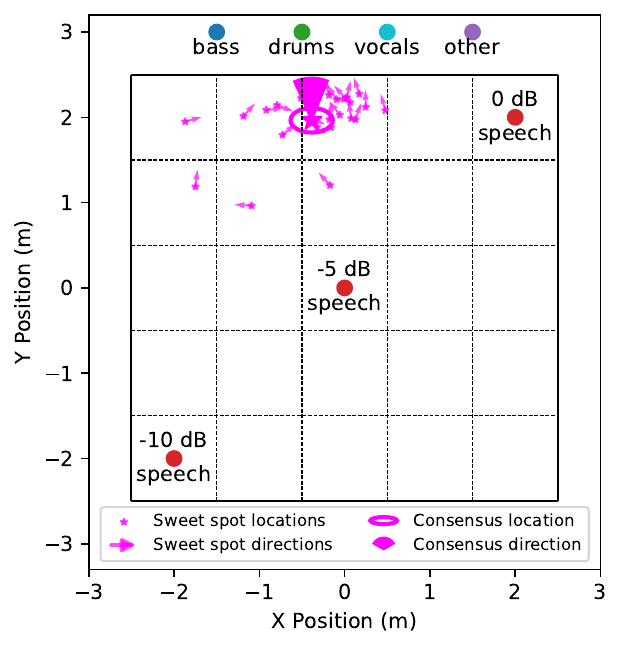}%
    \label{fig: sweetspots-s6-av}}
    \hfil
    \caption{Subjective sweet spot locations and directions across three audio-visual scenes with an increasing number of distractor sources, showing individual participant selections, aggregate means, and 95\% confidence intervals. Consensus sweet spots shift away from speech distractors while maintaining preferred acoustic balance relative to music sources.}
    \Description{This figure presents three spatial plots (a–c) showing subjective sweet spot locations and directions across audio-visual scenes S2-AV, S4-AV, and S6-AV with 1, 2, and 3 speech distractors respectively. Four target music sources sit at the top boundary. Across all distractor configurations, individual selections (pink stars) and mean consensus areas (magenta ellipses at Y ≈ 2m) remain tightly clustered near the frontal music stage, shifting slightly to avoid nearby speech distractors while maintaining direct orientation toward the music sources.}
    \label{fig: sweetspots}
\end{figure*}

Figure~\ref{fig: sweetspots} visualizes the subjective sweet spots across three audio-visual scenes featuring an increasing number of distractor sources. Individual participant sweet spot locations and directions are rendered as discrete markers and directional vectors, respectively. We also computed and plotted the 95\% confidence intervals for these distributions using spatial ellipses for locations and angular cones for orientations. The centroids and central axes of these confidence bounds represent the aggregate mean sweet spot location and direction across participants, while their spatial areas and angular spreads quantify the degree of inter-participant agreement.


The consensus sweet spot varied dynamically according to scene conditions. In general, sweet spot locations shifted away from distractor sources while maintaining a preferred balance of the music sources. Furthermore, the confidence ellipses exhibited greater lateral width than longitudinal depth, as spatial variance along the lateral axis ($x$-coordinate) was markedly higher than along the longitudinal axis ($y$-coordinate). This indicates that participants maintained a relatively consistent depth in front of the music sources, primarily making lateral adjustments to mitigate distractor interference and to maintain preferred musical balance. Meanwhile, individual and aggregate directional vectors demonstrate that participants consistently oriented toward the music sources across both auditory and audio-visual modalities.

\begin{figure*}[t]
    \centering
    \subfloat[S1 sweet spots]{\includegraphics[width=0.33\textwidth]{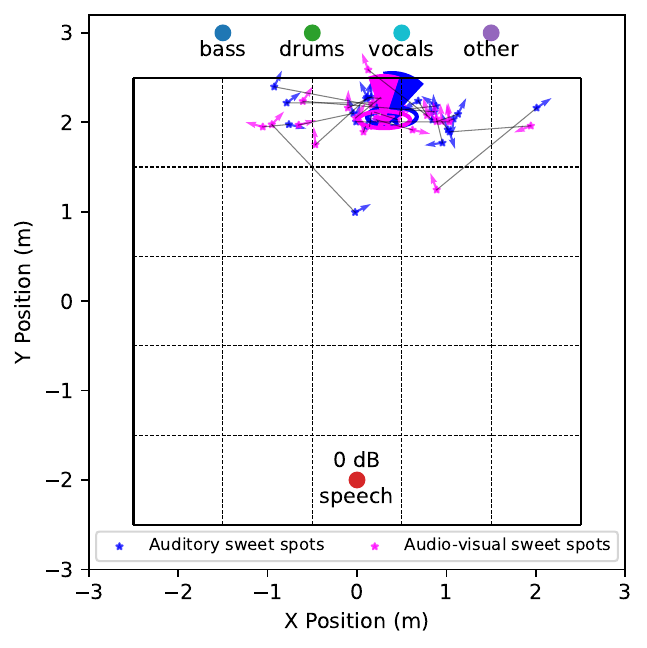}%
    \label{fig: cue-modalities-s1}}
    \hfil
    \subfloat[S3 sweet spots]{\includegraphics[width=0.33\textwidth]{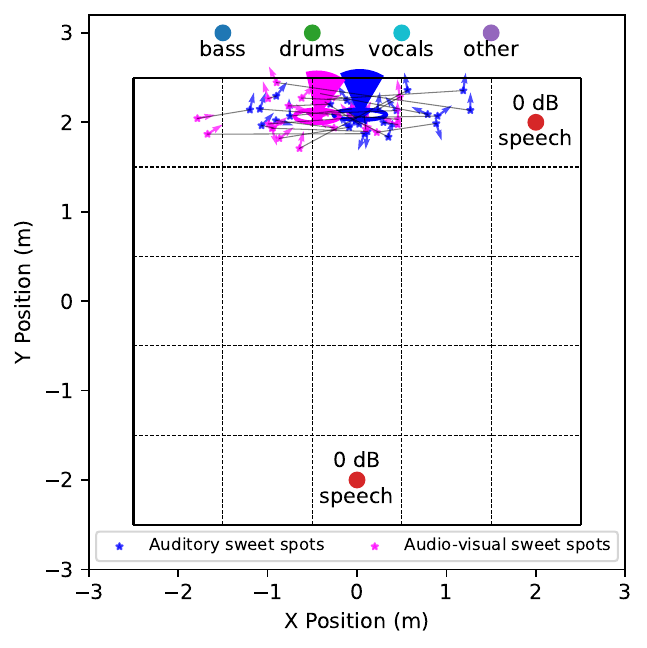}%
    \label{fig: cue-modalities-s3}}
    \subfloat[S5 sweet spots]{\includegraphics[width=0.33\textwidth]{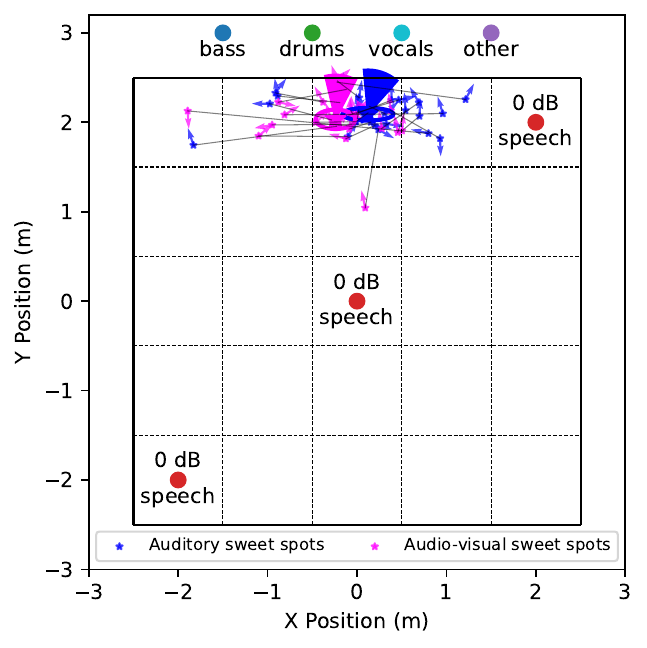}%
    \label{fig: cue-modalities-s5}}
    \hfil
    \caption{Subjective sweet spot locations and directions across auditory and audio-visual environments with increasing distractor density. Connecting lines highlight intra-participant shifts between modalities. Shifts become more pronounced as distractor sources increase, demonstrating that visual cues intensified participant avoidance from distractor sources.}
    \Description{This figure presents three 2D spatial plots (a–c) comparing subjective sweet spot selections between Auditory (blue) and Audio-Visual (pink) conditions across 1, 2, and 3 speech distractors (S1, S3, S5). Connecting gray lines show intra-participant shifts between modalities. Four target music sources sit at Y=3m. As distractor density increases from S1 to S5, the length of the connecting lines and the offset between blue and pink consensus ellipses increase, demonstrating that visual cues lead participants to shift further away from speech distractors.}
    \label{fig: cue-modalities}
\end{figure*}

Figure~\ref{fig: cue-modalities} illustrates the differences between subjective sweet spots obtained in auditory versus audio-visual environments, with connecting lines depicting intra-participant shifts across modalities. Notably, the 95\% confidence ellipses and cones exhibit incomplete spatial overlap, particularly in scenes with higher distractor density (Figures~\ref{fig: cue-modalities-s3} and \ref{fig: cue-modalities-s5}). This divergence demonstrates that integrating visual cues altered participant sweet spot selection.


Table~\ref{tab: participant-sweetspots} outlines the aggregate subjective sweet spot locations and directions across all evaluated scenes. Focusing first on spatial position, audio-visual environments shifted consensus sweet spot locations significantly further away from speech distractors relative to auditory conditions---as seen in Scene 3 (shifting leftward from $\bar{x} = 0.03$ in S3-A to $\bar{x} = -0.45$ in S3-AV; Figure~\ref{fig: cue-modalities-s3}) and Scene 5 (shifting from $\bar{x} = 0.12$ to $\bar{x} = -0.24$; Figure~\ref{fig: cue-modalities-s5}). This displacement suggests that visual representations of distractors heightened participant avoidance behavior. A two-way RM-ANOVA confirmed that both scene configuration ($p = 2.23 \times 10^{-9}$) and cue modality ($p = 1.38 \times 10^{-3}$) significantly affected lateral sweet spot positions ($x$-coordinates). Subsequent Tukey’s HSD post-hoc tests verified that audio-visual environments yielded sweet spots shifted significantly further left (away from distractor sources except in S1 and S4) compared to auditory conditions ($p = 6.89 \times 10^{-4}$). In contrast, neither scene configuration nor cue modality exerted a significant effect on longitudinal positions ($y$-coordinates), with mean viewing depths remaining tightly bounded across all scenes (ranging only from $\bar{y} = 1.97$ in S6-AV to $\bar{y} = 2.09$ in S4-A). These statistical findings confirm three primary spatial trends: (1) lateral sweet spot positions were dynamically adjusted according to distractor placement and relative music source balance; (2) longitudinal depth was maintained consistently in the region fronting the primary music sources; and (3) incorporating visual cues induced a significant lateral displacement away from distractor sources due to heightened distractor avoidance.


\begin{table}[t]
    \centering
    \caption{The subjective sweet spot locations and directions for each scene along with their 95\% confidence intervals. }
    \label{tab: participant-sweetspots}
    \begin{tabular}{lrrr|lrrr}
        \toprule
        Scene & $\bar{x} \pm \text{CI}_{0.95}$ & $\bar{y} \pm \text{CI}_{0.95}$ & $\bar{\phi} \pm \text{CI}_{0.95}$ & Scene & $\bar{x} \pm \text{CI}_{0.95}$ & $\bar{y} \pm \text{CI}_{0.95}$ & $\bar{\phi} \pm \text{CI}_{0.95}$ \\
        \midrule
        S1-A & $0.39 \pm 0.28$ & $2.06 \pm 0.11$ & $-17 \pm 26$ & S1-AV & $0.30 \pm 0.29$ & $2.03 \pm 0.09$ & $0 \pm 19$  \\
        S2-A & $0.11 \pm 0.32$ & $2.09 \pm 0.06$ & $-15 \pm 22$ & S2-AV & $-0.20 \pm 0.23$ & $1.99 \pm 0.13$ & $-2 \pm 22$ \\
        S3-A & $0.03 \pm 0.29$ & $2.08 \pm 0.06$ & $-3 \pm 27$ & S3-AV & $-0.45 \pm 0.25$ & $2.07 \pm 0.07$ & $-12 \pm 25$ \\
        S4-A & $0.71 \pm 0.31$ & $2.09 \pm 0.05$ & $-21 \pm 21$ & S4-AV & $0.55 \pm 0.32$ & $2.07 \pm 0.07$ & $8 \pm 24$ \\
        S5-A & $0.12 \pm 0.28$ & $2.09 \pm 0.07$ & $-18 \pm 23$ & S5-AV & $-0.24 \pm 0.22$ & $2.03 \pm 0.11$ & $-7 \pm 20$ \\
        S6-A & $-0.14 \pm 0.31$ & $2.09 \pm 0.05$ & $15 \pm 27$ & S6-AV & $-0.39 \pm 0.25$ & $1.97 \pm 0.14$ & $1 \pm 23$ \\
        \bottomrule
    \end{tabular}
\end{table}

Turning to listener directions, initial visual inspection of Table~\ref{tab: participant-sweetspots} suggested a shift in heading: consensus directions in auditory conditions generally faced front-right toward the vocal source, whereas audio-visual conditions realigned toward stage center—as seen in Scene 1 (from $\bar{\phi} = -17^\circ$ in S1-A to $0^\circ$ in S1-AV; Figure~\ref{fig: cue-modalities-s1}) and Scene 5 (from $\bar{\phi} = -18^\circ$ to $-7^\circ$; Figure~\ref{fig: cue-modalities-s5}). To formally evaluate these preferences, we conducted a two-way RM-ANOVA on angular data mapped to $[-180^\circ, 180^\circ]$ centered on forward facing ($\phi = 0^\circ$), satisfying RM-ANOVA normality assumptions due to tight clustering away from the $\pm 180^\circ$ phase discontinuity, i.e. most participant headings were toward the stage. The analysis revealed that heading directions were not significantly affected by scene configuration ($p = 0.150$) or cue modality ($p = 0.553$). Consequently, the initial hypothesis that visual cues systematically reoriented participants toward the stage center was not statistically supported, confirming that listener heading remains tightly aligned toward the stage regardless of environmental complexity or visual cues.

\subsection{Subjective and Objective Sweet Spot Maps}
\label{sec: results-sweetspotmaps}

This section addresses RQ4: How can objective quality metrics be evaluated against human-selected subjective sweet spots? To enable comparative similarity analyses against objective quality metrics, we generated subjective sweet spot maps by computing the normalized Mahalanobis scores as in Equation~\eqref{eqn: mahalanobis-scores} based on the participant sweet spot selections. Then, we produced the SNR and LQ objective sweet spot maps mirroring the subjective map computations, and observed how the objective predictions aligned with the subjective maps across scene configurations and modalities. 



\begin{figure*}[t]
    \centering
    \subfloat[S1-AV $\mathrm{SQ}_{\mathrm{ss}}$ map]{\includegraphics[width=0.33\textwidth]{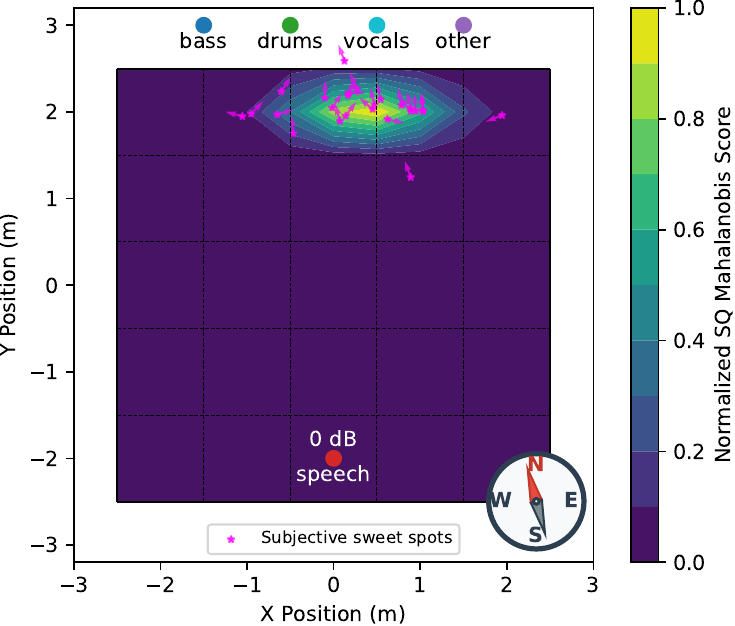}%
    \label{fig: sweetspotmap-subj-s1-av}}
    \hfil
    \subfloat[S1-AV $\mathrm{SNR}_{\mathrm{ss}}$ map]{\includegraphics[width=0.33\textwidth]{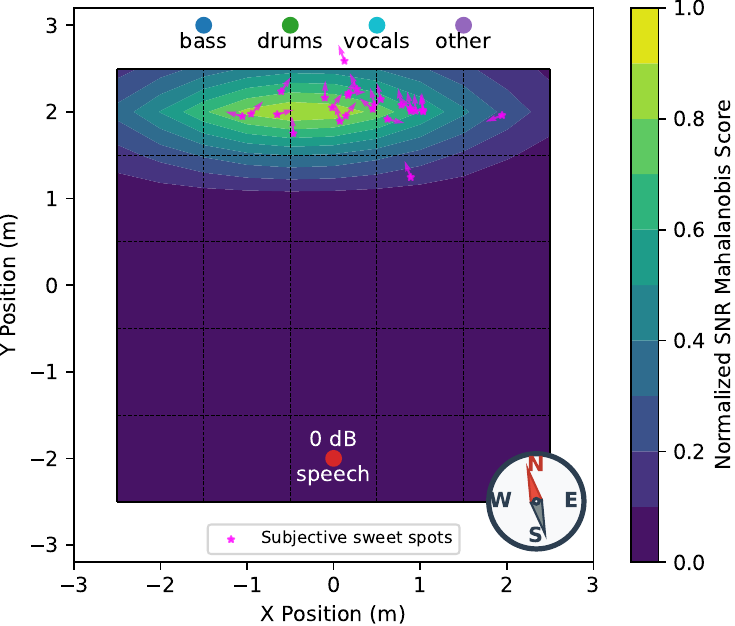}%
    \label{fig: sweetspotmap-snr-s1-av}}
    \subfloat[S1-AV $\mathrm{LQ}_{\mathrm{ss}}$ map]{\includegraphics[width=0.33\textwidth]{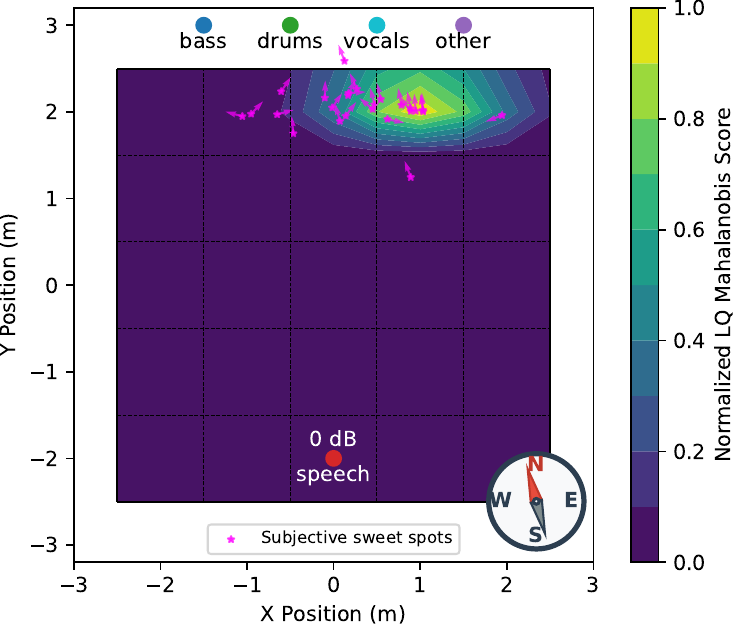}%
    \label{fig: sweetspotmap-lq-s1-av}}
    \hfil
    \break
    \subfloat[S4-A $\mathrm{SQ}_{\mathrm{ss}}$ map]{\includegraphics[width=0.33\textwidth]{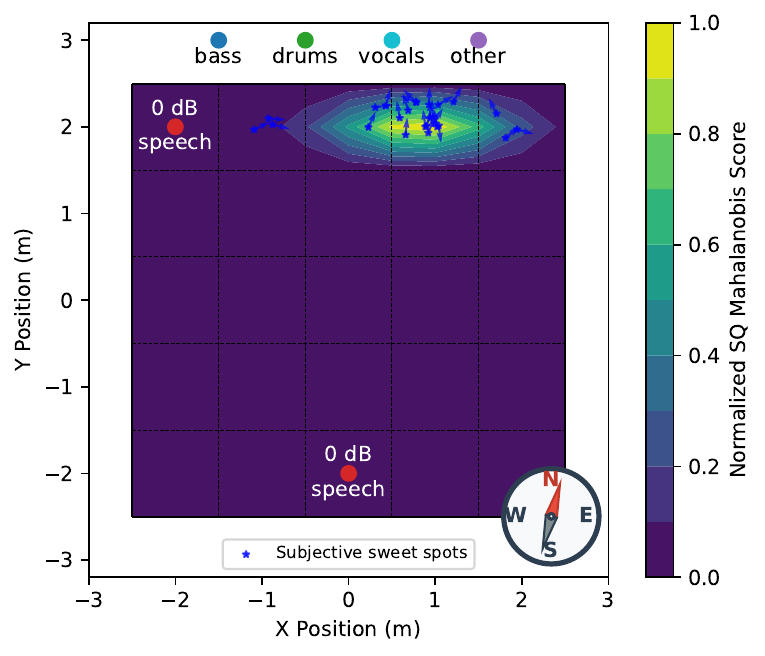}%
    \label{fig: sweetspotmap-subj-s4-a}}
    \hfil
    \subfloat[S4-A $\mathrm{SNR}_{\mathrm{ss}}$ map]{\includegraphics[width=0.33\textwidth]{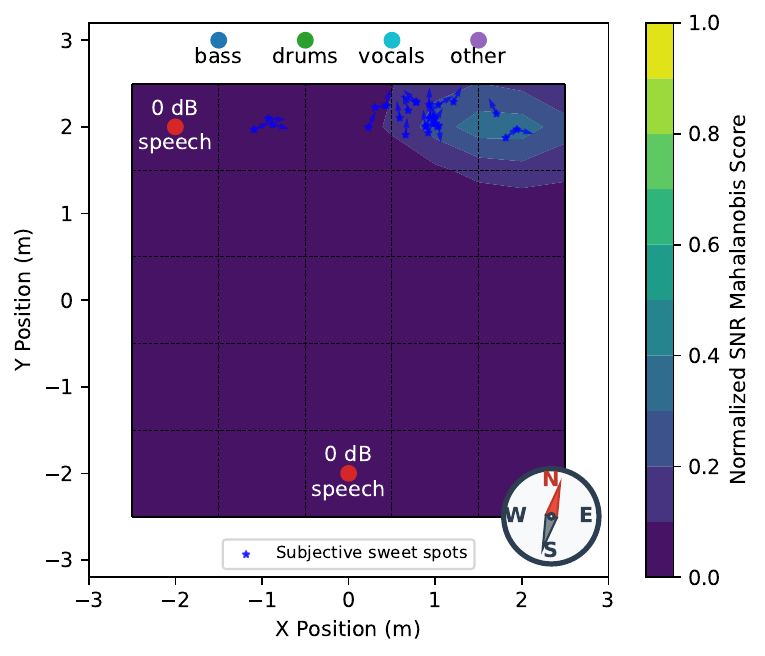}%
    \label{fig: sweetspotmap-snr-s4-a}}
    \subfloat[S4-A $\mathrm{LQ}_{\mathrm{ss}}$ map]{\includegraphics[width=0.33\textwidth]{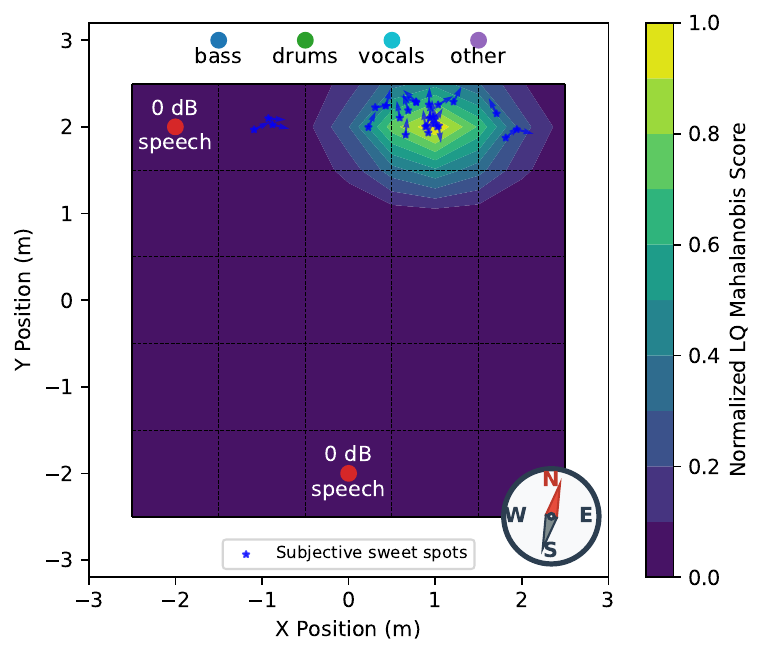}%
    \label{fig: sweetspotmap-lq-s4-a}}
    \hfil
    \caption{The subjective (first column), SNR (second column), and LQ (third column) sweet spot maps of an audio-visual (top) and an auditory scene (bottom), along with their subjective sweet spots. The compass for each scene indicates the direction with the maximum subjective score. We can observe that the LQ maps had more similar sweet spot patterns with the subjective maps compared to the SNR maps. Meanwhile, the SNR maps had sweet spot patterns that were either laterally spread out (S1) or were barely visible (S4) as the sweet spot was found at a different direction compared to the consensus optimal listening direction.}
    \Description{This figure presents six 2D spatial sweet spot maps in a 2x3 grid comparing subjective quality, SNR, and LQ sweet spot scores from 0.0 (purple) to 1.0 (yellow) across an audio-visual scene S1-AV (top row) and an auditory scene S4-A (bottom row). Target music sources sit at Y=3m, and speech distractors are marked with red dots. Individual subjective sweet spot choices appear as pink and blue stars for audio-visual and auditory scenes respectively. Across both environments, LQ maps (c, f) yield tight, concentrated high-score peaks that match subjective sweet spot distributions (a, d) significantly better than SNR maps (b, e), which appear either laterally diffuse or attenuated}
    \label{fig: sweetspot-maps}
\end{figure*}

Figures~\ref{fig: sweetspotmap-subj-s1-av} and~\ref{fig: sweetspotmap-subj-s4-a} depict representative subjective sweet spot maps for audio-visual (S1-AV) and auditory (S4-A) scenes, respectively. Because these contour maps vary dynamically with listener direction, they are affixed at the fixed heading angle corresponding to the global maximum subjective quality score $\mathrm{SQ}_{\mathrm{ss}}(\mathbf{x})$. High-density regions (colored brightly) denote locations with the highest probability of being selected as a sweet spot. The complete dynamic shift of these sweet spots across varying listener directions is demonstrated in the supplemental animations from the GitHub repository.

The generated maps effectively capture empirical sweet spot distributions, with high-density peaks containing the sweet spot selections from the participants. Because these subjective maps were constructed strictly from locations rated as "excellent," unselected regions exhibit near-zero values. These near-zero regions do not imply absolute poor quality, but rather indicate areas that participants bypassed during navigation in search of the optimal quality. In contrast, raw objective quality maps (Figs.~\ref{fig: naviqual-results-contour-1} and~\ref{fig: naviqual-results-contour-2}) predict continuous quality ratings spanning the entire scale from "bad" to "excellent." To establish a direct basis for comparison, objective quality maps were transformed into objective sweet spot maps following the methodology in Section~\ref{sec: methods-correlating}. Specifically, objective sweet spots were extracted using a quality threshold of $T = 0.8$, corresponding to a score exceeding 4 out of 5 ("excellent") on a five-point scale.


Figures~\ref{fig: sweetspotmap-snr-s1-av}--\ref{fig: sweetspotmap-lq-s4-a} compare the resulting SNR- and LQ-derived objective sweet spot maps against subjective benchmarks for S1 and S4 at their respective consensus directions. In Scene 1 (audio-visual; top row of Fig.~\ref{fig: sweetspot-maps}), the SNR metric predicted a laterally diffuse sweet spot in front near the stage. Conversely, the LQ metric predicted a tightly localized sweet spot focused directly in front of the vocal source, closely matching the subjective sweet spot. Looking at the sweet spot animations, rotational dynamics reveal distinct behaviors between metrics: as the listener direction rotates counter-clockwise, the LQ-predicted sweet spot shifts rightward to maintain participant heading toward the music sources---which closely mirrors the subjective map behavior. In contrast, the SNR-predicted sweet spot shifts leftward, orienting away from the speech distractor rather than toward primary music sources.

In Scene 4 (auditory; bottom row of Fig.~\ref{fig: sweetspot-maps}), SNR-based sweet spots were virtually absent at the consensus direction because optimal SNR occurred off-axis (oriented eastward facing away from distractors). Conversely, the LQ metric accurately localized the sweet spot at the forward-facing consensus direction, yielding spatial distributions and rotational shift patterns that coincided closely with the subjective benchmark. Collectively, these comparative evaluations demonstrate that the LQ metric serves as a significantly more reliable predictor of subjective sweet spot locations and orientation dynamics than conventional SNR.

\begin{figure}[t]
    \centering
    \subfloat[S2 polar plots]{\includegraphics[width=0.33\textwidth]{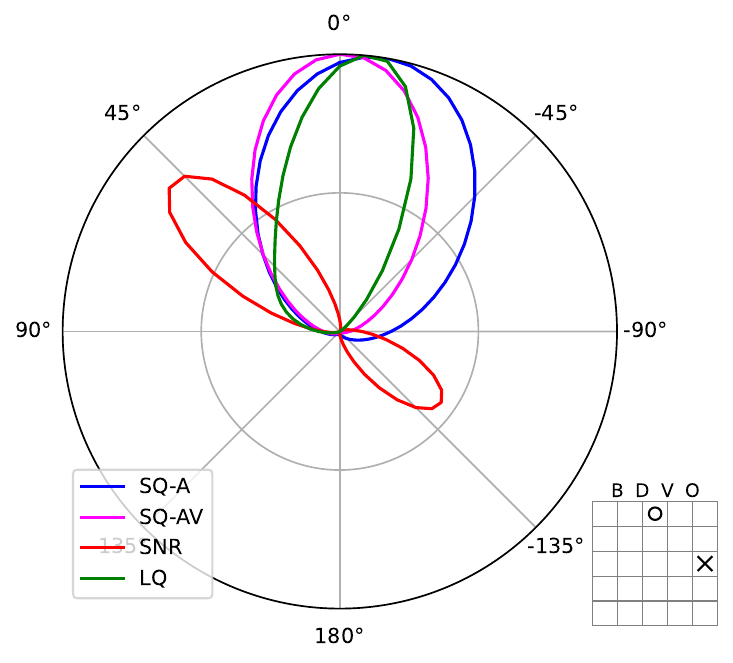}%
    \label{fig: polar-s2}}
    \subfloat[S4 polar plots]{\includegraphics[width=0.33\textwidth]{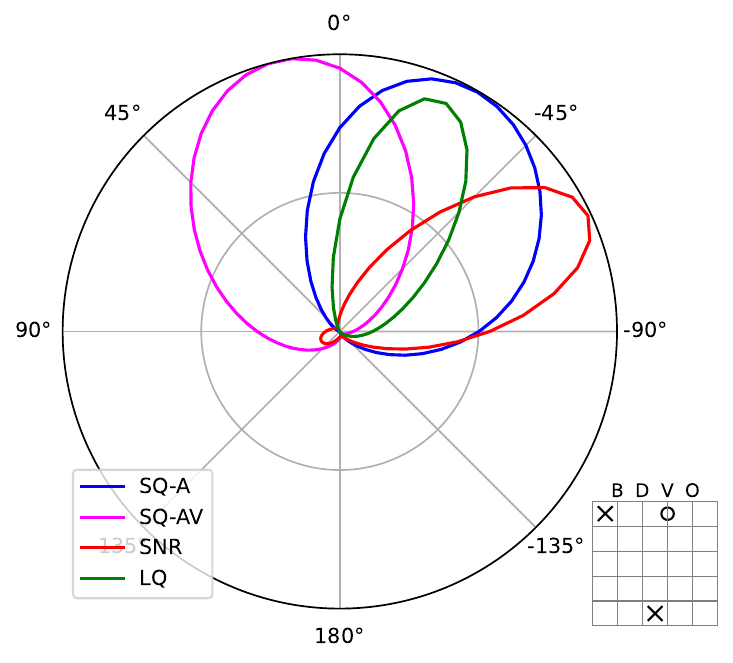}%
    \label{fig: polar-s4}}
    \subfloat[S5 polar plots]{\includegraphics[width=0.33\textwidth]{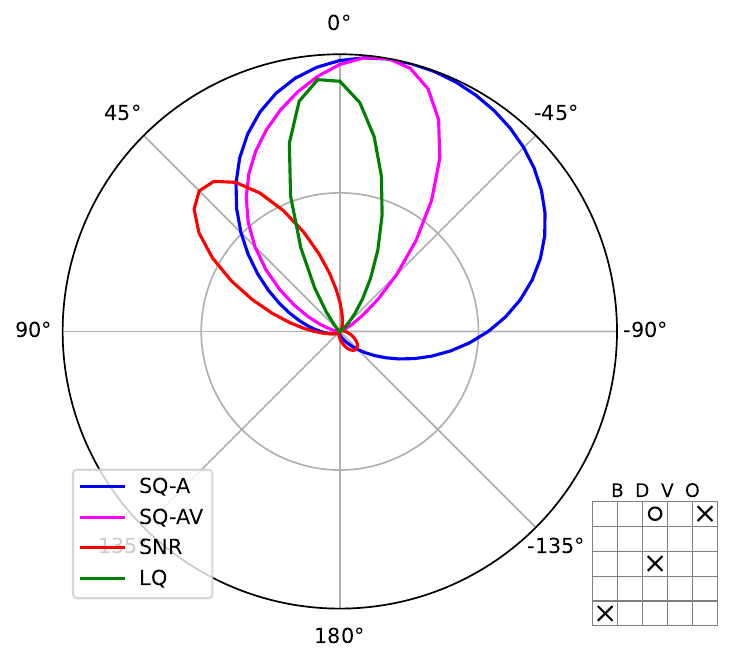}%
    \label{fig: polar-s5}}
    \hfil
    \caption{The polar plots of the normalized subjective and objective sweet spot scores in scenes S2, S4, and S5 in both auditory and audio-visual conditions. They were produced based on a fixed subjective sweet spot location roughly shown in the lower right of these plots. We can observe that the subjective sweet spot directions (\textit{blue} and \textit{magenta}) were better predicted by the LQ polar plots (\textit{green)} compared to SNR (\textit{red}). Furthermore, we can observe that the SNR sweet spot scores do not reach the maximum value at the fixed location in S5, since the SNR sweet spot location was located further to the left away from the distractor sources.}
    \Description{This figure presents three side-by-side polar plots (a–c) comparing directional audio quality scores at fixed locations in scenes S2, S4, and S5. The plots display normalized subjective quality for auditory (blue) and audio-visual (magenta) conditions against objective LQ (green) and SNR (red) metrics. Across all scenes, the LQ metric (green) closely matches the front-facing directional lobes of the subjective scores (blue and magenta) around 0 degrees, whereas the SNR metric (red) predicts off-axis directions or attenuated scores.}
    \label{fig: polar_plots}
\end{figure}

To evaluate directional preferences, Figure~\ref{fig: polar_plots} illustrates orientation profiles of subjective and objective sweet spot scores at fixed spatial coordinates $(x,y)$, selected where the audio-visual subjective quality score $\mathrm{SQ}_{\mathrm{ss}}(\mathbf{x})$ reached its global maximum. These polar plots contrast objective predictions (red and green curves) against subjective orientation scores in auditory (blue) and audio-visual (magenta) conditions, with spatial scene schematics inset in the lower-right corner of each plot.

In Scene 4 (Fig.~\ref{fig: polar-s4}), the SNR metric predicted an optimal heading directed eastward, pointing directly away from speech distractors rather than toward the music sources. Conversely, the LQ metric predicted a north-east heading—--balancing orientation toward the music sources with distractor avoidance--—which coincided closely with the auditory consensus direction (while audio-visual consensus displayed a slight inward shift toward stage center).

Similarly, in Scenes 2 and 5 (Figs.~\ref{fig: polar-s2} and~\ref{fig: polar-s5}), SNR incorrectly predicted optimal headings directed westward away from distractors; furthermore, SNR scores failed to reach maximum threshold at this reference coordinate since its optimal value was located further to the left. In contrast, the LQ metric consistently predicted headings oriented toward the music sources despite speech distractors positioned to the right or rear, mirroring both auditory and audio-visual subjective benchmarks.

Overall, the LQ metric predicted the empirical listening direction preferences more accurately compared to SNR. Fundamentally, the superior predictive accuracy of the LQ metric stems from its joint perceptual formulation: whereas conventional SNR evaluates listening quality purely by penalizing distractor interference, LQ jointly optimizes for relative music source balance and the perceptual threshold of distractor annoyance.

\subsection{Similarity between the Subjective and Objective Maps}
\label{sec: results-similarity}

This section addresses two research questions: How accurately do objective metrics predict empirical subjective sweet spots (RQ5), and how robust are objective sweet spot predictions across variations in musical genre, distractor configurations, and visual cues (RQ6)? To quantitatively evaluate predictive performance, we computed the similarity metrics between objective and subjective sweet spot maps across a $11 \times 11 \times 72$ spatial-directional grid (comprising $8\,712$ unique $(x, y, \phi)$ points), and compared the scores across six scene set ups in auditory and audio-visual modalities. Table~\ref{tab: similarity-metric-scores} summarizes these similarity scores across all evaluated scenes and modalities. Bold values indicate the objective metric yielding superior prediction accuracy, while underlined values denote the modality achieving higher similarity. Aggregate mean scores across all scenes are provided to summarize overall performance.


\begin{table*}[t]
    \centering
    \caption{Similarity metric scores between the subjective and objective sweet spot maps}
    \label{tab: similarity-metric-scores}
    \begin{tabular}{lcccccccccccc}
        \toprule
        & \multicolumn{4}{c}{MAE} & \multicolumn{4}{c}{SSIM} & \multicolumn{4}{c}{EMD} \\
        \cmidrule(lr){2-5} \cmidrule(lr){6-9} \cmidrule(lr){10-13}
        & \multicolumn{2}{c}{Auditory} & \multicolumn{2}{c}{Audio-visual} & \multicolumn{2}{c}{Auditory} & \multicolumn{2}{c}{Audio-visual} & \multicolumn{2}{c}{Auditory} & \multicolumn{2}{c}{Audio-visual} \\
        \cmidrule(lr){2-3} \cmidrule(lr){4-5} \cmidrule(lr){6-7} \cmidrule(lr){8-9} \cmidrule(lr){10-11} \cmidrule(lr){12-13}     
        Scene & SNR & LQ & SNR & LQ & SNR & LQ & SNR & LQ & SNR & LQ & SNR & LQ \\
        \midrule
        S1 & 0.0275 & \textbf{0.0127}       & 0.0297 & {\ul \textbf{0.0101}} & 0.7609 & \textbf{0.8467}       & 0.7251 & {\ul \textbf{0.9179}} & 0.1941 & \textbf{0.1645}       & 0.1904 & {\ul \textbf{0.1367}} \\
        S2 & 0.0236 & \textbf{0.0138}       & 0.0214 & {\ul \textbf{0.0119}} & 0.8082 & \textbf{0.8735}       & 0.8099 & {\ul \textbf{0.8838}} & 0.2182 & {\ul \textbf{0.1499}} & 0.2104 & \textbf{0.1537}       \\
        S3 & 0.0231 & \textbf{0.0179}       & 0.0185 & {\ul \textbf{0.0135}} & 0.7797 & \textbf{0.8094}       & 0.8372 & {\ul \textbf{0.8772}} & 0.2327 & \textbf{0.1862}       & 0.2027 & {\ul \textbf{0.1545}} \\
        S4 & 0.0163 & {\ul \textbf{0.0103}} & 0.0216 & \textbf{0.0137}       & 0.8425 & {\ul \textbf{0.9144}} & 0.8187 & \textbf{0.8628}       & 0.1576 & {\ul \textbf{0.1345}} & 0.2014 & \textbf{0.1709}       \\
        S5 & 0.0206 & \textbf{0.0145}       & 0.0169 & {\ul \textbf{0.0100}} & 0.8390 & \textbf{0.8751}       & 0.8436 & {\ul \textbf{0.9222}} & 0.2307 & \textbf{0.1781}       & 0.2053 & {\ul \textbf{0.1523}} \\
        S6 & 0.0225 & \textbf{0.0158}       & 0.0220 & {\ul \textbf{0.0117}} & 0.8387 & \textbf{0.8716}       & 0.8486 & {\ul \textbf{0.8971}} & 0.2401 & \textbf{0.1985}       & 0.2216 & {\ul \textbf{0.1566}} \\
        \midrule
        Mean & $0.0223$ & $\mathbf{0.0142}$ & $0.0217$ & $\underline{\mathbf{0.0118}}$ & $0.8115$ & $\mathbf{0.8651}$ & $0.8139$ & $\underline{\mathbf{0.8935}}$ & $0.2122$ & $\mathbf{0.1686}$ & $0.2053$ & $\underline{\mathbf{0.1541}}$ \\
        \bottomrule
    \end{tabular}
\end{table*}

Across all six scenes and both cue modalities, the LQ metric consistently outperformed SNR in predicting the subjective sweet spots. Superior performance was maintained across all three evaluation dimensions: absolute value error (MAE), spatial structure and gradient patterns (SSIM), and spatial-directional probability distributions (EMD). These findings demonstrate that advanced binaural quality metrics are essential for accurately modeling spatial audio quality in AR environments, where human quality perception relies primarily on interaural phase degradations rather than simple signal intensity degradations.


Furthermore, objective metrics predicted subjective sweet spots more accurately in audio-visual environments than in auditory environments in four of the six scenes. As illustrated in the polar plots of Scenes 2 and 5 (Figs.~\ref{fig: polar-s2} and \ref{fig: polar-s5}), introducing visual cues produced more concentrated distributions toward stage center, in contrast to auditory-only conditions where the listener headings were more spread out, slightly penalizing the similarity scores. Conversely, in Scene 4 (Fig.~\ref{fig: polar-s4}), the visual cues pulled listener orientations inward toward stage center, whereas the consensus direction in auditory conditions was directed northeast similar to the LQ predictions, resulting in lower similarity scores.


On aggregate, objective quality maps exhibited higher predictive similarity to subjective maps in audio-visual conditions than in auditory-only conditions, even though the maps were derived strictly from binaural acoustic signals without visual parameter inputs. This finding highlights a key perceptual insight: visual cues provide vital spatial grounding in AR environments, reducing navigation uncertainty and guiding participants toward sweet spots that align more closely with the optimal binaural acoustic quality.


\section{Conclusion}
\noindent

In this work, we investigated listener navigation behavior and subjective perceptual experience in AR live music environments, pairing empirical observations with objective quality modeling to identify listener sweet spots. We introduced an extensive and novel 6-DoF subjective evaluation framework backed with quantitative analysis relating listener navigation data with objective quality metrics. We also introduced a novel approach for assessing objective metric performance on 6-DoF subjective data by computing similarity metrics against subjective quality maps. 


Our empirical findings reveal three primary spatial and behavioral insights: First, audio-visual conditions enabled direct and efficient trajectories, whereas auditory-only conditions prompted exploratory navigation paths, demonstrating the value of providing visual cues for efficient navigation toward the sweet spots (RQ1). Second, while participants consistently selected sweet spots in the forward region facing the stage, lateral positioning varied according to individual preferences of both musical source balance and distractor avoidance (RQ2). Third, introducing visual cues shifted the subjective sweet spots further away from speech distractors, suggesting that the added visual information subtly affects the subjective experience (RQ3). Collectively, we established that listeners form consensus sweet spots according to the scene conditions through our subjective framework.


By constructing subjective sweet spot maps from empirical distributions (RQ4), we established a quantitative benchmark to evaluate objective quality metrics. Comparative similarity analyses revealed that Localization Quality (LQ) predicted the subjective sweet spots more accurately than conventional Signal-to-Noise Ratio (SNR) in both auditory and audio-visual modalities (RQ5). Furthermore, these objective predictions demonstrated strong robustness across musical genres, distractor configurations (up to three distractors), distractor intensities, and cue modalities (RQ6). Collectively, we established that advanced binaural quality metrics measuring phase degradations are well-suited for modeling spatial audio sweet spots in AR compared to conventional intensity-based metrics.


While this study validates NAVIQUAL-XR across diverse scene configurations, the experimental scope was deliberately bounded to anechoic acoustic conditions in AR. This controlled baseline was necessary to first isolate direct-sound perceptual dynamics and determine whether objective quality metrics reliably predict human sweet spot selections without confounding room reflections. Extending this framework to complex reverberant environments constitutes a primary direction for future work. Towards this goal, preliminary pilot evaluations comparing objective quality maps under anechoic and reverberant conditions are detailed in Appendix~\ref{sec: reverberant}, confirming strong correlation between anechoic and reverberant maps for both objective metrics ($R\ge 0.9841$). Because these pilot evaluations demonstrate that objective quality maps are not significantly altered despite reverberation, we hypothesize that empirical human perception will exhibit a similar trend in reverberant AR environments.

The NAVIQUAL-XR framework advances spatial audio methodology by bridging the gap between stationary 3-DoF testing and dynamic 6-DoF interactive environments. By unifying empirical participant navigation with objective quality maps, NAVIQUAL-XR provides a foundation for designing immersive AR environments aligned with human subjective experience. Beyond general live music applications, this framework establishes a pathway toward real-time, adaptive AR listening experiences tailored for all individuals including the hard-of-hearing.

\begin{acks}
This work was funded by the European Commission under the Marie Skłodowska-Curie Actions postdoctoral fellowship grant ID 101151676 (ARM4CI) and by Taighde Éireann – Research Ireland through the Rinn Artificial Intelligence (Rinn AI) centre. Ethical approval for the subjective tests was obtained from the UCD Human Research Ethics Committee. 
\end{acks}

\bibliographystyle{ACM-Reference-Format}
\bibliography{sample-base}

\appendix

\section{Magic Leap 2 Spatializer System}
\label{sec: ml2-spatializer}

In our study (Figure~\ref{fig: naviqual-xr-graphic}), subjective tests with participants were conducted using Magic Leap 2 (ML2) AR glasses, which employ a proprietary spatializer (Soundfield Audio) to render the AR environment. Conversely, the NAVIQUAL objective quality maps \cite{Tolentino2026naviqual} were generated using a distinct head-related transfer function---specifically, the KU100 model from the open-source SADIE II database \cite{Armstrong2018}. To confirm the validity of evaluating ML2 subjective responses against NAVIQUAL objective maps despite the spatializer mismatch, we empirically verified the NAVIQUAL predictions against a representative subset of physical ML2 binaural recordings.

Using Unity 6, we captured the rendered binaural audio signals---comprising the target music signal $S_{\text{music}}(t)$ and speech distractor(s) $S_{\text{speech}}(t)$---across the test environments (Table~\ref{tab: scene-setup}) using the ML2 Soundfield Audio spatializer engine. Binaural signals were captured across a $\qty{5}{\meter} \times \qty{5}{\meter}$ listening area at a spatial resolution of $\qty{1}{\meter}$ (25 locations) for four listener directions ($\phi \in \{0^{\circ}, 90^{\circ}, 180^{\circ}, 270^{\circ}\}$), yielding 100 binaural recordings per scene. Each measuring point required approximately $\qty{15}{\second}$ to capture both 7-second binaural signals required for computing the objective metric, highlighting a significant computational bottleneck when producing the objective quality maps of six AR scenes having 8,712 recording points for each. From the representative subset of ML2 recordings, Signal-to-Noise Ratio (SNR) and Localization Quality (LQ) metrics were calculated and compared directly against NAVIQUAL objective quality maps at matching locations, directions, and scenes.

Table~\ref{tab: correlation-spatializers} presents the Pearson correlation coefficients ($R$) between the ML2-rendered binaural metrics and the NAVIQUAL objective metrics. Correlations were evaluated across 25 spatial locations independently for each direction slice, ensuring that the full dynamic range of SNR and LQ was spanned within each environment. The results demonstrate strong linear agreement for both SNR ($R \ge 0.8779$) and LQ ($R \ge 0.9175$), robust against diverse scene configurations and musical genres. This confirms that the NAVIQUAL objective quality predictions serve as a reliable proxy for analyzing the subjective listener evaluations of the ML2-rendered AR environments.  

\begin{table*}[t]
    \centering
    \caption{Pearson correlation coefficients $R$ between the objective metrics outputs of NAVIQUAL and Magic Leap 2 recordings.}
    \label{tab: correlation-spatializers}
    \begin{tabular}{lcccccccccc}
        \toprule
        & \multicolumn{2}{c}{$\phi = 0^{\circ}$} & \multicolumn{2}{c}{$\phi = 90^{\circ}$} & \multicolumn{2}{c}{$\phi = 180^{\circ}$} & \multicolumn{2}{c}{$\phi = 270^{\circ}$} & \multicolumn{2}{c}{Mean} \\
        \cmidrule(lr){2-3} \cmidrule(lr){4-5} \cmidrule(lr){6-7} \cmidrule(lr){8-9} \cmidrule(lr){10-11}
        Scene & SNR & LQ & SNR & LQ & SNR & LQ & SNR & LQ & SNR & LQ \\
        \midrule
        S1 & 0.9332 & 0.9362 & 0.9290 & 0.9311 & 0.9448 & 0.9390 & 0.9203 & 0.9175 & \textbf{0.9318} & \textbf{0.9310} \\
        S2 & 0.9911 & 0.9954 & 0.9964 & 0.9931 & 0.9955 & 0.9904 & 0.9871 & 0.9856 & \textbf{0.9925} & \textbf{0.9911} \\
        S3 & 0.8784 & 0.9175 & 0.9697 & 0.9624 & 0.9003 & 0.9335 & 0.9613 & 0.9466 & \textbf{0.9274} & \textbf{0.9400} \\
        S4 & 0.8779 & 0.9427 & 0.9660 & 0.9549 & 0.8961 & 0.9314 & 0.9497 & 0.9375 & \textbf{0.9224} & \textbf{0.9416} \\
        S5 & 0.9849 & 0.9751 & 0.9914 & 0.9816 & 0.9777 & 0.9564 & 0.9881 & 0.9858 & \textbf{0.9855} & \textbf{0.9747} \\
        S6 & 0.9717 & 0.9769 & 0.9876 & 0.9695 & 0.9693 & 0.9368 & 0.9869 & 0.9675 & \textbf{0.9789} & \textbf{0.9604} \\

        \bottomrule
    \end{tabular}
\end{table*}

\section{NAVIQUAL in Reverberant Conditions}
\label{sec: reverberant}

In our work, we conducted our subjective tests with the Magic Leap 2 (ML2) glasses featuring its proprietary Soundfield Audio spatializer plugin to render the AR live music environments. We deliberately limited our scope into anechoic conditions to validate our objective model under diverse scene configurations without reverberation. We acknowledge that in real scenarios, live music environments are inherently reverberant based on the room acoustic properties. Thus, we conducted preliminary investigations on how introducing reverberation into our virtual environments affect the objective quality metrics, namely the signal-to-noise ratio (SNR) and the localization quality (LQ) derived from BINAQUAL\cite{Shariat2026binaqual}.

Our virtual live music environments consisting of multiple music and speech sound sources were synthesized using Binamix \cite{Barry2025binamix}. The tool renders binaural signals using a collection of HRTF models from the SADIE II database \cite{Armstrong2018}, recorded under anechoic (head-related impulse responses, HRIR) and reverberant conditions (binaural room impulse responses, BRIR). The KU100 HRIR model was used to synthesize the virtual environments defined in Table~\ref{tab: scene-setup}, on which the the objective quality maps were produced using NAVIQUAL \cite{Tolentino2026naviqual}. To verify whether reverberation affected the objective maps, the same virtual environments were rendered using the KU100 BRIR model. Notably, the HRIR model contained 8802 recording points from the database, whereas the BRIR model only contained 50 points. The lower sampling resolution from the BRIR model was compensated by synthesizing the binaural signals based on the nearest neighboring point from the database, thus preserving the actual BRIR recordings without introducing spectral artifacts from interpolating multiple BRIR points. Binaural signals were recorded from representative locations and directions with a sampling density of $\qty{0.5}{\meter}$ translational and $45^{\circ}$ rotational resolutions, resulting in $968$ data points. The Pearson correlation coefficients between the objective quality metrics produced from the HRIR synthesis and BRIR synthesis were computed across the scenes to verify whether reverberation significantly affected the metrics.

\begin{figure*}[t]
    \centering
    \subfloat[S2 anechoic LQ map]{\includegraphics[width=0.33\textwidth]{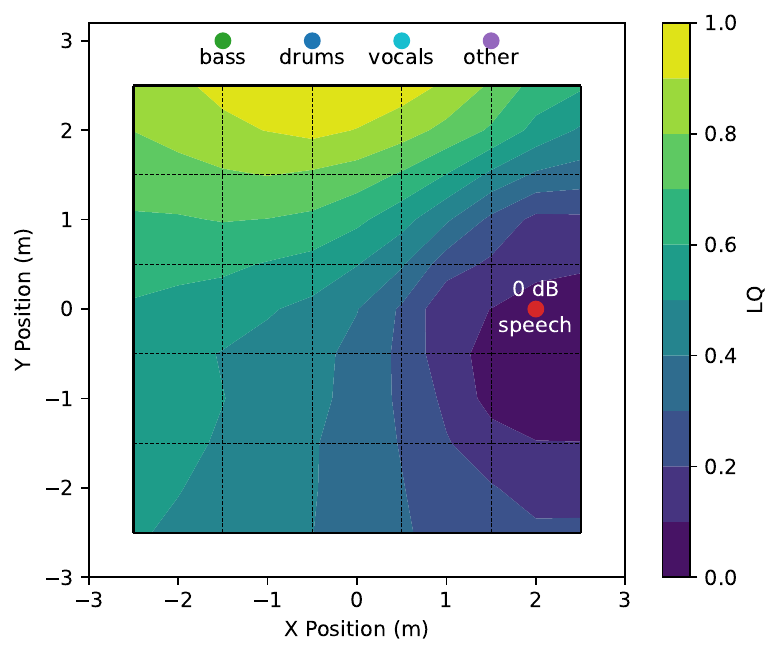}%
    \label{fig: lqmap-s2-hrir}}
    \hfil
    \subfloat[S4 anechoic LQ map]{\includegraphics[width=0.33\textwidth]{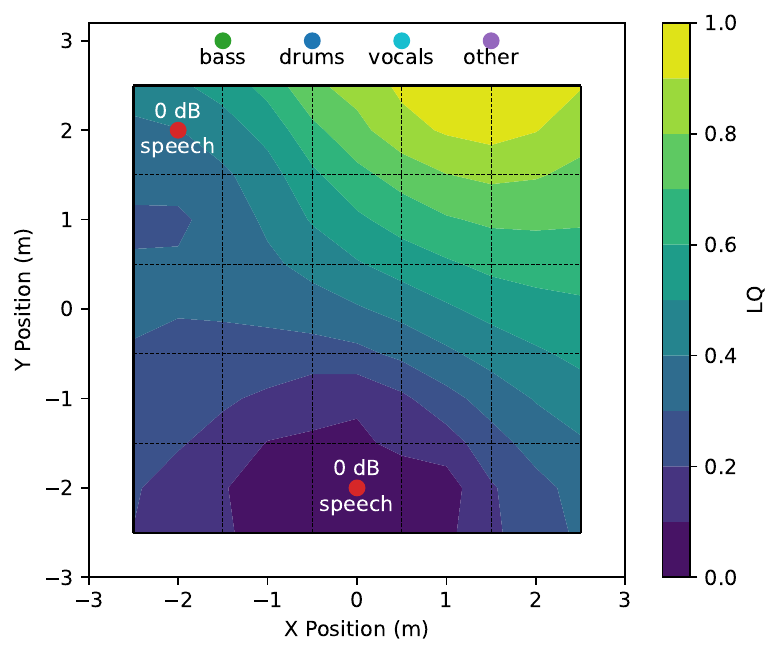}%
    \label{fig: lqmap-s4-hrir}}
    \subfloat[S6 anechoic LQ map]{\includegraphics[width=0.33\textwidth]{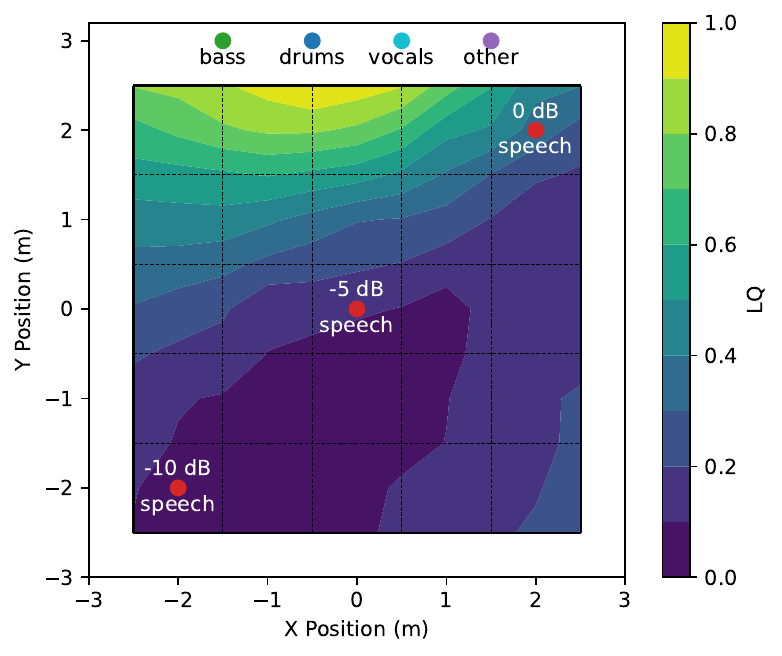}%
    \label{fig: lqmap-s6-hrir}}
    \hfil
    \break
    \subfloat[S2 reverberant LQ map]{\includegraphics[width=0.33\textwidth]{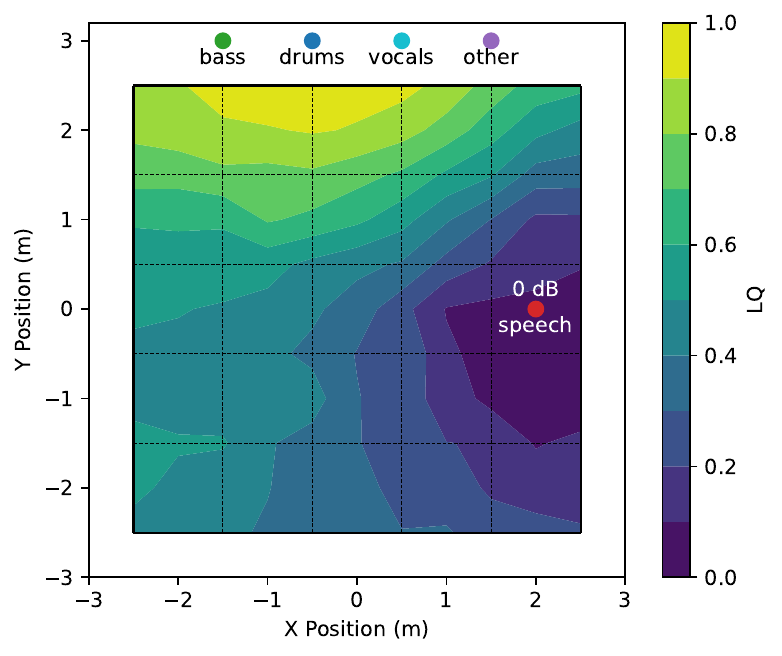}%
    \label{fig: lqmap-s2-brir}}
    \hfil
    \subfloat[S4 reverberant LQ map]{\includegraphics[width=0.33\textwidth]{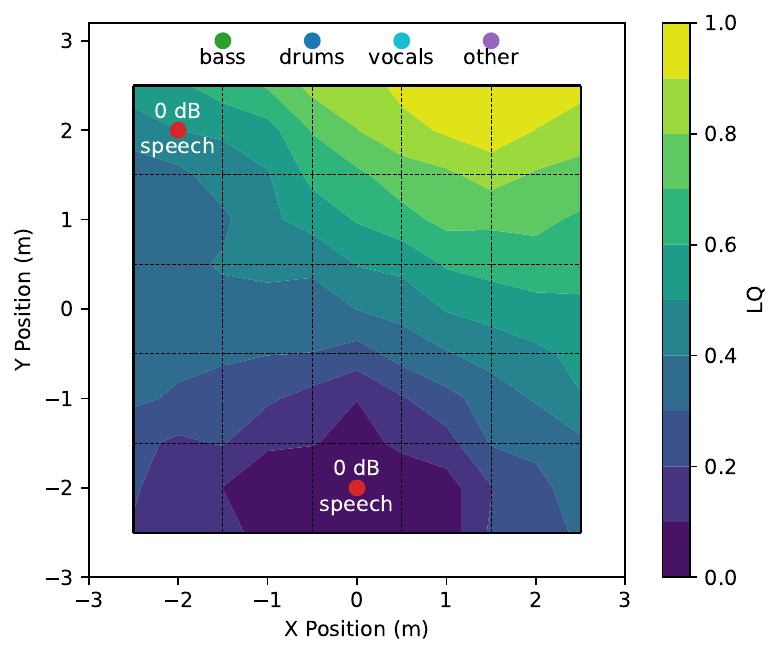}%
    \label{fig: lqmap-s4-brir}}
    \subfloat[S6 reverberant LQ map]{\includegraphics[width=0.33\textwidth]{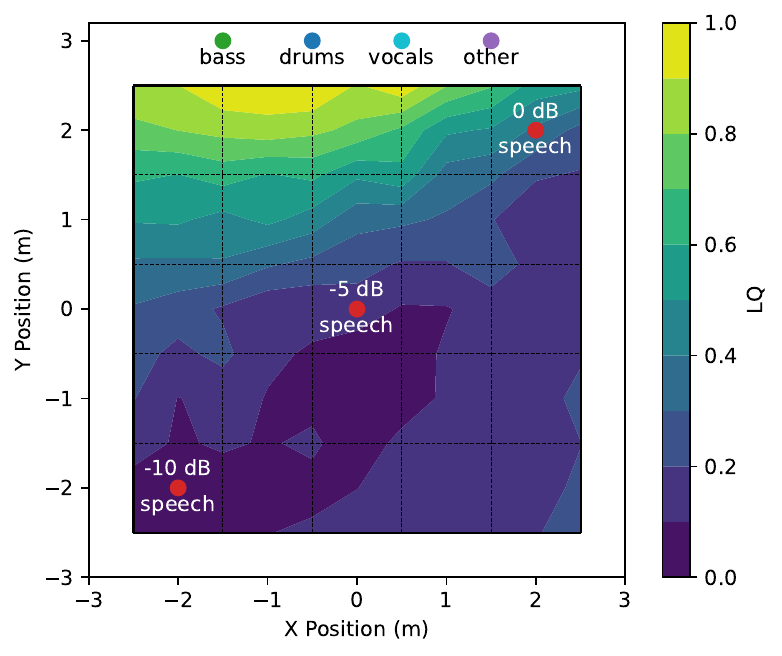}%
    \label{fig: lqmap-s6-brir}}
    \hfil

    \caption{The objective quality maps based on LQ of three scenes under anechoic (top) and reverberant (bottom) conditions. The objective quality contour levels are not significantly affected despite reverberation. }
    \Description{This figure presents six 2D spatial contour maps in a 2x3 grid comparing normalized LQ scores from 0.0 (purple) to 1.0 (yellow) across scenes S2, S4, and S6 under anechoic (top row a–c) and reverberant (bottom row d–f) conditions. Four music sources sit at the frontal stage, while red speech distractor dots are placed at varying locations. Across all three scene configurations, the presence of room reverberation does not significantly alter the objective LQ contour levels or spatial distributions compared to their anechoic counterparts.}
    \label{fig: naviqual-reverberant}
\end{figure*}

Figure~\ref{fig: naviqual-reverberant} visualizes the LQ objective quality maps of three distinct scenes under anechoic (top) and reveberant (bottom) conditions in the forward-facing direction $\phi = 0$. We can observe subtle variations on the objective quality maps under reverberant conditions. This can be attributed by the lower data point resolution from the BRIR model, as the contour lines separating the objective quality regions exhibited more discontinuous patterns compared to the anechoic conditions. Nonetheless, the objective quality regions were preserved in scene S2 and S4, while we can observe a slight westward shift of the objective sweet spot in scene S6.

Table~\ref{tab: correlation-reverberant} presents the correlation coefficients between the SNR and LQ objective quality metrics under anechoic (HRIR) and reverberant (BRIR) conditions. Near-perfect linear agreement on both SNR ($R \ge 0.9824$) and LQ ($R \ge 0.9843$) objective metrics was observed between the anechoic and reverberant environments, signifying that adding reverberation does not significantly alter the objective metric patterns across listener locations and directions. We established in our experiments (Section~\ref{sec: results-sweetspotmaps} and Section~\ref{sec: results-similarity}) that the objective quality metrics accurately predicted the subjective sweet spots of listeners immersed in anechoic AR environments. Thus, we hypothesize that objective metric predictions will maintain high similarity with the subjective sweet spots in reverberant environments, which can be validated in future work. 

\begin{table*}[t]
    \centering
    \caption{Pearson correlation coefficients $R$ between objective metrics outputs under anechoic and reverberant conditions.}
    \label{tab: correlation-reverberant}
    \begin{tabular}{lcccccc}
        \toprule
        Metric & Scene 1 & Scene 2 & Scene 3 & Scene 4 & Scene 5 & Scene 6 \\
        \midrule
        SNR & 0.9946 & 0.9884 & 0.9938 & 0.9904 & 0.9876 & 0.9894 \\
        LQ & 0.9943 & 0.9896 & 0.9881 & 0.9877 & 0.9860 & 0.9841 \\
        \bottomrule
    \end{tabular}
\end{table*}


\end{document}